\documentclass[12pt]{article}

\usepackage{xcolor}
\usepackage{amssymb}
\usepackage{subcaption}
\usepackage{amsmath,amsfonts}
\usepackage{graphicx}
\usepackage{caption}
\usepackage{geometry}
\usepackage{booktabs}
\usepackage{setspace}

\title{Five-Gene Expression Formula Accurately Detects Hepatocellular Carcinoma Tumors}
\author{Aram Ansary Ogholbake$^1$ and Qiang Cheng$^{1,\ast}$\\
\\
\small $^1$Institute for Biomedical Informatics, Department of Computer Science,\\
\small University of Kentucky, Lexington, KY, United States \\
\small \textbf{Corresponding author:} Qiang Cheng, Email: qianq.cheng@uky.edu}
\date{}
\begin{document}
\maketitle

\begin{abstract}
Hepatocellular carcinoma (HCC) is one of the leading causes of cancer-related deaths worldwide. Several diagnostic methods, such as imaging modalities and Serum Alpha-Fetoprotein (AFP) testing, have been used for HCC detection; however, their effectiveness is limited to later stages of the disease. In contrast, transcriptomic analysis of biposy samples has shown promise for early detection. While machine learning techniques have been applied to transcriptomic data for cancer detection, their clinical adoption remains limited due to challenges such as poor generalizability across different datasets, lack of interpretability, and high computational complexity. To address these limitations, we developed a novel predictive formula for HCC detection using the Kolmogorov-Arnold Network (KAN). This formula is based on the expression levels of five genes: VIPR1, CYP1A2, FCN3, ECM1, and LIFR. Derived from the GSE25097 dataset, the formula offers a simple, interpretable, efficient, and accessible approach for HCC identification. It achieves 99\% accuracy on the GSE25097 test set and demonstrates robust performance on six additional independent datasets, achieving accuracies of above 90\% in all cases. These findings highlight the critical role of these five genes as biomarkers for HCC detection, offering a foundation for future research and clinical applications to improve HCC diagnostic approaches.
\end{abstract}

\noindent \textbf{Keywords:} Hepatocellular carcinoma, Predictive formula, Gene expression, Neural network

\section{Introduction}

 Liver cancer is the fourth leading cause of cancer-related deaths worldwide \cite{siegel2021cancer}.  Hepatocellular carcinoma (HCC), the most common type of primary liver cancer, accounts for approximately 90\% of all liver tumors. With its incidence projected to rise significantly in the coming decade \cite{badwei2023hepatocellular}, detecting the cancer at an early stage is crucial for improving patient outcomes. There are several methods available to detect HCC, including Serum Alpha-Fetoprotein (AFP), imaging modalities such as Ultrasound (US), Computed Tomography (CT), Magnetic Resonance Imaging (MRI), as well as liver biopsy and molecular analysis. However, each method has some limitations. AFP testing alone has limited sensitivity and specificity in early detection of HCC \cite{parikh2020biomarkers,bruix2011management}. Imaging techniques, while useful, also have shown limitations. For instance, significant heterogeneity in their sensitivity, and their suboptimal specificity necessitates further tests, complicating 
the diagnostic process \cite{tzartzeva2018surveillance,konerman2019frequency,atiq2017assessment}. Moreover, imaging modalities such as US are operator dependent, and some lesions are difficult to access \cite{wong2015elevated}. 
 Although, CT and MRI generally have better performance, they are most effective at later stages of HCC, typically when tumor size is larger than 2 cm in diameter. Furthermore, the costs of these modalities are high and are used when an evident signal appears \cite{chan2024biomarkers}. Additionally, the AFP levels may not be elevated in early stage HCC, and combining AFP testing with imaging techniques may still provide false negative result \cite{galle2019biology}. Liver biopsy may also lead to false negative results, particularly if samples are obtained from the wrong location \cite{villanueva2010hepatocellular,el2011current,zhang2024development}. However, the molecular analysis of biopsy samples, particularly through transcriptomic approaches, has shown potential to overcome these limitations and improving the HCC detection.  

Transcriptomics provides a thorough molecular analysis of biological systems, enabling the prediction of disease outcomes by analyzing changes associated with disease progression \cite{zhang2024integration}. High-throughput technologies, such as microarrays, facilitate this process by simultaneous assessment of tens of thousands of gene expression levels, helping to identify existing patterns correlated to the disease prognosis. Recent studies have utilized machine learning on gene expression data to improve cancer detection for various types, including breast cancer \cite{mohamed2023bio,yaqoob2024improving,coleto2022multi}, colorectal cancer \cite{amniouel2024high,ahmadieh2023using}, kidney cancer \cite{li2022identification}, ovarian cancer \cite{vaiyapuri2022red}, prostate cancer \cite{rostami2022gene}, and HCC \cite{zhang2024development}. 
Although machine learning techniques show promise for cancer detection, they carry a risk of overfitting to training data, potentially limiting their ability to generalize to new, unseen data. Moreover, high computational resource demands and the need for specialized expertise limit their practical application, making it challenging for clinicians to integrate them in everyday practice. In \cite{zhang2024development}, which uses machine learning-based predictors for classifying HCC tissues and cirrhosis tissues without HCC (CwoHCC), the classification process involved at least 18 secreted genes. This is a relatively high number, as not all of these genes might be available in different datasets. Furthermore, although the study reported a high sensitivity rate using the MRMD+SVM predictor, it did not report specificity results. Specificity is crucial because misclassifying normal samples as cancerous could have serious consequences for healthy individuals. 

In this study, we present a formula using a recent model called Kolmogrove-Arnold Network (KAN) \cite{liu2024kan} trained on GSE25097 to detect HCC. This formula is simple to use, requiring no machine learning expertise and relying solely on the expression levels of five genes:  VIPR1, CYP1A2, FCN3, ECM1 and LIFR. The transparency and interpretability of our model aim to build trust among clinician, bridging the gap between advanced computational techniques and clinical applications. The formula allows clinicians to observe the role of each gene in the HCC identification process. Despite its simplicity, our formula shows great performance. On the GSE25097 test set, our formula shows high sensitivity and specificity and achieves an accuracy of 99\%. When tested across six independent datasets— GSE60502, GSE57957, GSE64041, GSE121248, GSE47197, and GSE76297-the formula achieves great performance using different metrics. While previous research has identified VIPR1, CYP1A2, FCN3, ECM1 and LIFR individually as downregulated in HCC \cite{fu2022activation,yu2021cyp1a2,ma2023fcn3,chen2016ecm1,luo2015lifr}, these genes have not been explored together as a combined set of biomarkers for tumor prediction ( {the role of each individual gene in HCC can be found in Subsection 3.3)}.  Moreover, based on differential expression analysis, these genes are not among the top significant genes and in some datasets, they are found to be not significant at all. Our study demonstrates that focusing on these biomarkers collectively can achieve very high predictive accuracy on HCC, offering a novel perspective on their role in HCC detection and highlighting their potential as powerful tools for early diagnosis. In summary, we developed a predictive formula for differentiating normal from HCC cancerous tissues at different stages and validated its performance across multiple datasets. We analyzed the expression levels of VIPR1, CYP1A2, FCN3, ECM1 and LIFR in both normal and cancerous samples, providing comparative results between the two groups. We also provided associated drugs with the five genes, offering insights into potential therapeutic options. Further analyses, including differential gene expression and genes correlations heat maps, supported the biological significance of our findings.

\section{Meterial and methods}
\subsection{Overview}

\begin{figure*}[]
\centering
\includegraphics[width=0.83\textwidth]{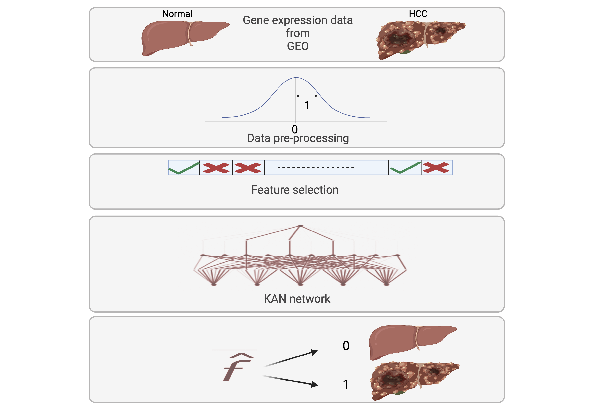}
\caption{Overview of the proposed method for HCC tumor classification.}
\label{fig:overview}
\end{figure*}

We define the dataset as follows: \[
\mathcal{D} = \{(x_i, y_i) \mid i = 1, \dots, n\},
\]

\noindent where:
\begin{itemize}
    \item $x_i \in \mathbb{R}^m$ represents the gene expression profile of the $i$-th sample, consisting of $m$ genes.
    \item $y_i \in \{0, 1\}$  denotes the label for the $i$-th sample, where $y_i = 1$ indicates that the sample $i$ has HCC and $y_i = 0$ indicates that the sample is normal. 
\end{itemize}
Our goal is to develop a predictive formula using the training data to  classify samples in the test data as either normal or HCC. This process involves multiple steps, including dataset preparation, data pre-processing, feature selection, and modeling with KAN (Figure \ref{fig:overview}).

\subsection{Datasets}
We utilized GSE25097 dataset for developing the predictive formula for HCC detection. This dataset consists of 268 HCC tumor samples and 243 adjacent non-tumorous samples from early-to-advanced stages. The dataset was randomly divided into a training set and a test set in an 80:20 ratio.

To validate the robustness of the predictive formula, we further tested it on six additional independent datasets: GSE60502, GSE57957, GSE64041, GSE121248, GSE47197, and GSE76297. { The detail of each dataset is provided below.\\
\textbf{GSE60502: } Includes 36 paired HCC and non-tumorous liver tissues, collected from surgically resected livers during HCC treatment.\\
\textbf{GSE57957: } Comprises 39 HCC tumor tissues and 39 matched adjacent non-tumorous liver tissues, all obtained through surgical resection.\\
\textbf{GSE64041: } Includes 60 liver biopsies from HCC patients across all tumor stages, with one biopsy taken from the tumor and another from the non-tumorous liver for each patient.\\
\textbf{GSE121248:} Contains 70 chronic hepatitis B induced HCC tumor tissue samples and 37 normal liver tissue samples.\\
\textbf{GSE47197:} Features 124 liver tissue samples, of which 61 are hepatitis B virus induced HCC and 63 are normal liver tissues. \\
\textbf{GSE76297:} Includes liver tissue samples from patients with HCC and Cholangiocarcinoma (CCA). For this study, we focused exclusively on the 62 HCC tumor tissues and 60 non-tumorous liver tissues, excluding CCA samples.}

These datasets allowed us to verify the predictive formula's generalizability performance.

\subsection{Data Pre-processing }

We employed gene based z-score normalization to calculate the standardized values in GSE25097 dataset as:
\begin{equation}
z_{ij} = \frac{x_{ij} - \mu_{j}}{\sigma_{j}},
\end{equation}
\noindent where $z_{ij}$ is normalized expression value for gene $j$ in sample $i$, $x_{ij}$ is the original expression value for gene $j$ in sample $i$. $\mu_{j}$ is the mean expression value of gene $j$ across all samples and $\sigma_{j}$ is the standard deviation of gene $j$ across all samples. Specifically, for each gene, its mean expression value was subtracted from its observed values, and the result was divided by the standard deviation of that gene across all samples. This allows for fair comparisons across genes and minimizing biases caused by varying scales in gene expression. 

\subsection{Feature Selection } 

To have an effective classification, efficient feature selection model is needed. Specifically, a small size datasets with high dimensionallity often contain noisy features, which can significantly impact a model's classification performance. Hence, it's important to find informational set of features to train the model. We used the Fisher-Markov selector \cite{cheng2010fisher} which is shown to be effective to identify the most useful features in describing essential differences among different groups in high dimensional data. 

Unlike many other feature selection methods that either struggle in handling high dimensional data or are limited to find local optimum, the Fisher-Markov selector adopts a systematic approach to optimize for sparsity and discriminativeness in high-dimensional settings. It leverages Markov random field optimization techniques to solve the formulated objective functions, enabling efficient feature selection. This makes it suitable for our classification purpose which uses high dimensional data.

In the Fisher-Markov selector method, there is only one free parameter denoted by $\gamma$. We set its value to -0.5 as suggested in the paper. This selector found the top 5 genes which are:  VIPR1, CYP1A2, FCN3, ECM1 and LIFR. The GitHub code is available here: https://github.com/aramansary/Fisher-Markov-Selector.

\begin{figure}[h]
\centering
\includegraphics[width=0.45\textwidth]{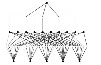}
\caption{Structure of the KAN network used to derive the predictive formula.}
\label{fig:network}
\end{figure}

\subsection{Modeling with KAN }
To obtain the predictive formula for HCC detection, we used KAN which is introduced as a promising alternative to Multi-Layer Perceptrons (MLPs), recently.  KANs are inspired by the Kolmogrov-Arnold representation theorem which states that any multivariate function can be expressed as a finite composition of univariate functions and sums. Unlike traditional neural networks which use fixed activation functions on nodes, KANs use learnable activation functions on edges, parametrized as splines. Therefore, KANs have no linear weight matrices, and each weight parameter is replaced by a learnable 1D function. This leads to greater interpretability while capturing complex relationships between features.

 The main principle of KAN lies in approximating a target function $f(z)$ as a composition of univariate functions and sums, represented by:

\begin{equation}
    f(z) = f(z_1 \cdots z_{m'}) = \sum_{q=1}^{k} \Phi_q \left( \sum_{p=1}^{m'} \phi_{q,p}(z_p) \right),
\end{equation}

\noindent where $\phi_{q,p}$ and $\Phi_q$ are univariate functions learned by the network. k is the number of univariate functions and p is number of features used after feature selection stage.

The formulation can be written in matrix form and it can generalize to deeper and wider architectures, allowing for smooth and accurate modeling of multivariate functions. 
This is done by stacking multiple layers, represented as:

\begin{equation}
KAN(z) = (\Phi_L \circ \Phi_{L-1} \circ \cdots \circ \Phi_1)(z),
\end{equation}

\noindent where $\Phi_L$ represents the learnable function matrix at layer $L$ as:
\begin{equation}
   \mathbf{\Phi} =
\begin{pmatrix}
\phi_{1,1}(\cdot) & \cdots & \phi_{1,n_{\text{in}}}(\cdot) \\
\vdots & \ddots & \vdots \\
\phi_{n_{\text{out}},1}(\cdot) & \cdots & \phi_{n_{\text{out}},n_{\text{in}}}(\cdot)
\end{pmatrix} .
\end{equation}

\noindent Here, $n_{\text{in}}$ represents the number of inputs and $n_{\text{out}}$ represents the number of outputs.

KAN was chosen for its shown high accuracy and great interpretability, making it well-suited for modeling complex gene expression data. By leveraging spline-based univariate functions and compositional structures, KAN effectively can capture nonlinear relationship in gene expression data, leading to great approximation and ensuring accurate HCC detection. Additionally, KAN provides an interface to define activation functions in a specified symbolic form. This feature, combined with iterative grid search for optimizing affine parameters, ensures the model achieves an optimal fit to the data. The final derived symbolic formula offers a clear and interpretable representation of how each gene contributes to the prediction which may enhance trust in real-world diagnostic settings.

We implemented a fully-connected KAN with L=2 layers  
\begin{equation}
KAN(z) = (\Phi_2 \circ \Phi_{1})(z),
\end{equation}

\noindent where $\Phi_{1}$ has the five selected genes found in the feature selection part as input features and it has 10 output features ($n_{\text{in}}$=5 and $n_{\text{out}}$=10). $\Phi_{2}$ then has 10 input features and 1 output for binary classification ($n_{\text{in}}$=10 and $n_{\text{out}}$=1). Figure \ref{fig:network} demonstrates the architecture of the KAN model we used. To enhance the efficiency of the model, we applied sparsity regularization during training. Specifically, we used the L1 norm of the activation function, coupled with entropy regularization. Overall penalty strength was set to 0.01 and entropy penalty strength was set to 1. We trained the model using 20 epochs with a built-in LBFGS optimization procedure which is known to be efficient in handling small to medium-sized datasets. 

After training, we set the symbolic functions for activations. As suggested in \cite{liu2024kan}, we used Sympy to compute the formula of the output node. This resulted in an explicit closed-form formula for HCC classification, enabling both interpretability and ease of application.

\begin{table}[]
\centering
\begin{tabular}{|l|c|c|c|c|}
\hline
\textbf{Dataset} & \textbf{Precision} & \textbf{Recall (Sensitivity)} & \textbf{Specificity} & \textbf{F1 Score} \\ \hline
\textbf{GSE25097 (test set)}  & 0.98 & 1.00 & 0.98 & 0.99 \\ \hline
\textbf{GSE60502 (full data)} & 1.00 & 0.94 & 1.00 & 0.97 \\ \hline
\textbf{GSE57957 (full data)}  & 0.90 & 0.92 & 0.90 & 0.91 \\ \hline
\textbf{GSE64041 (full data)}  & 0.98 & 0.83 & 0.98 & 0.90 \\ \hline
\textbf{GSE121248 (full data)} & 0.98 & 0.89 & 0.97 & 0.93 \\ \hline
\textbf{GSE47197 (full data)}  & 0.96 & 0.85 & 0.97 & 0.90 \\ \hline
\textbf{GSE76297 (full data)}  & 0.98 & 0.94 & 0.98 & 0.96 \\ \hline
\end{tabular}

\caption{Performance of our formula in different metrics for various datasets.}
\label{tab:metrics}
\end{table}

\subsection{Derived Formula for HCC}
Our computed formula for HCC classification is:
\begin{multline}
\hat{f} = -1.21 \cdot \sin\Big( 
    0.46 \cdot \sin(0.48 \cdot \text{VIPR1} - 4.15) + 
    0.19 \cdot \sin(0.75 \cdot \text{FCN3} + 2.17) + \\
    0.77 \cdot \sin(0.48 \cdot \text{ECM1} + 2.04) - 
    0.70 \cdot \sin(0.4 \cdot \text{LIFR} - 7.38) - \\
    0.03 \cdot \tanh(10.0 \cdot \text{CYP1A2} - 10.0) - 6.91 
\Big) - \\
3.0 \cdot \tan\Big(
    0.03 \cdot (0.5 \cdot \text{VIPR1} + 1)^2 - 
    0.12 \cdot \sin(0.33 \cdot \text{FCN3} + 8.34) - \\
    0.15 \cdot \sin(0.53 \cdot \text{ECM1} + 8.38) + 
    0.01 \cdot \tanh(8.6 \cdot \text{CYP1A2} - 10.0) + \\
    0.10 \cdot \tanh(0.61 \cdot \text{LIFR} - 1.06) - 6.89
\Big) - 2.07
\label{eq:formula}
\end{multline}
\noindent where $VIPR1$, $CYP1A2$, $FCN3$, $ECM1$ and $LIFR$ are normalized expressions of the corresponding genes. 

To apply the formula for HCC classification, the following steps should be followed:
\begin{itemize}
    \item Perform z-score normalization on the expression values of the five genes across all samples of the dataset.
    \item Plug the normalized expression values into Equation (\ref{eq:formula}).
    \item Classify the sample according to the calculated value of $\hat{f}$:   If 
$\hat{f} < 0.5$, classify as healthy tissue;  If $\hat{f} \geq 0.5$, classify as HCC tissue. 
\end{itemize}

\begin{table}[]
\centering
\begin{tabular}{|l|c|c|c|c|c|}
\hline
\textbf{Dataset} & \textbf{KNN} & \textbf{RF} & \textbf{GB} & \textbf{SVM} & \textbf{Our Method} \\ \hline
\textbf{GSE60502}   & \textbf{97.22} & 66.67 & 86.11 & 50.00 & \textbf{97.22} \\ 
\textbf{GSE57957}   & \textbf{93.59} & 53.85 & 50.00 & 52.56 & 92.00 \\ 
\textbf{GSE64041}   & 88.33          & 71.67 & 81.67 & 50.00 & \textbf{90.83} \\ 
\textbf{GSE121248}  & \textbf{94.39} & 85.05 & 70.09 & 65.42 & 91.59 \\ 
\textbf{GSE47197}   & 77.42          & 80.65 & 79.03 & 80.65 & \textbf{91.13} \\ 
\textbf{GSE76297}   & 94.21          & 71.07 & 91.74 & 57.02 & \textbf{95.87} \\ \hline
\textbf{Average Performance} & 90.86 & 71.49 & 76.44 & 59.27 & \textbf{93.11} \\ 
\hline
\end{tabular}

\caption{Comparison of our method to 4 different classifiers, all of which are trained on the dataset GSE25097, across 6 independent datasets.}
\label{tab:average_performance}
\end{table}

\section{Results}
\subsection{Performance Evaluation of the Predictive Formula}

Table \ref{tab:metrics} provides a detailed breakdown of the performance metrics for the predictive formula across seven independent datasets. This table includes 4 different metrics: precision, sensitivity, specificity, and F1 score. 

Precision is the ratio of true predicted cancerous samples over the total number of the predicted cancerous samples:
\begin{equation}
    Precison = \frac{TP}{TP+FP}
\end{equation}
Sensitivity is the ratio of true predicted cancerous samples over the total number of the cancerous samples:
\begin{equation}
    Sensitivity = \frac{TP}{TP+FN}
\end{equation}
Specificity is the ratio of true predicted healthy samples over the total number of the healthy samples:
\begin{equation}
    Specificity = \frac{TN}{TN+FP}
\end{equation}
F1 score is calculated as:
\begin{equation}
    F1 = \frac{2*Precision*Recall}{Precision+Recall}
\end{equation}
{F1 score combines both precision and recall into a single, balanced measure. This helps evaluate how well a model performs overall, especially when the data may be imbalanced.} 
The results demonstrate a great performance of our formula on the GSE25097 test set. On  this dataset, the formula achieves perfect sensitivity, 98\% specificity, 98\% precision, and 99\% F1 score, highlighting its robust capability to accurately identify HCC cancerous tissues.

The performance on the GSE60502 dataset, remains impressive but shows some decline in the F1 score. The formula achieves a sensitivity of 94\% indicating a high true positive rate, and a specificity of 100\%, reflecting its ability to correctly identify non-tumor tissues. The precision of 100\% and F1 score of 97\% further emphasize the formula's effectiveness.

Similarly, on the datasets GSE57957, GSE64041 GSE121248, GSE47197, and GSE76297, the formula consistently achieves F1 scores exceeding 90\%, with strong precision, sensitivity, and specificity, demonstrating its reliability across diverse datasets.

We have also provided Receiver Operating Characteristic (ROC) curves and confusion matrices (Supplementary Figure S1 and  Figure S2) to demonstrate the predictive power of the formula across different datasets. The confusion matrix visualizes the distribution of true positives, true negatives, false positives, and false negatives, offering insights into the formula's performance.

In the GSE25097 dataset, the results indicate an almost perfect Area Under Curve (AUC) of 1, with the confusion matrix revealing 1 false positives and no false negatives. For the GSE60502 dataset, only 1 sample was misclassified, comprising 1 false negatives  resulting in an AUC of 0.98. In the GSE57957 dataset, the AUC is 0.94, with three false negatives and four false positives. The GSE64041 has 1 false positive and 10 false negatives, with an AUC of 0.95. The GSE121248 dataset achieves an AUC of 0.96, with one false positive and eight false negatives. The GSE47197 dataset attains an AUC of 0.93, with two false positives and nine false negatives. Lastly, the GSE76297 achieves an AUC of 0.98 with 1 false positive and four false negatives.
\begin{figure*}[]
\centering
\begin{subfigure}[b]{\textwidth}
\centering
\includegraphics[width=\textwidth]{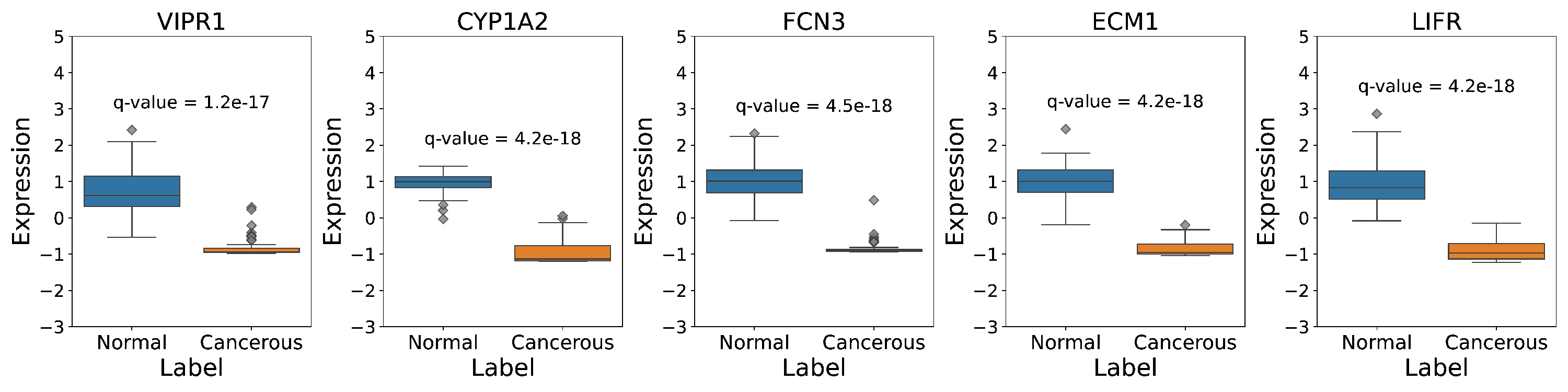}
\put(-430,115){\textbf{A}} 
\caption{}
\end{subfigure}
 \hfill
 \begin{subfigure}[b]{\textwidth}
\centering
\includegraphics[width=\textwidth]{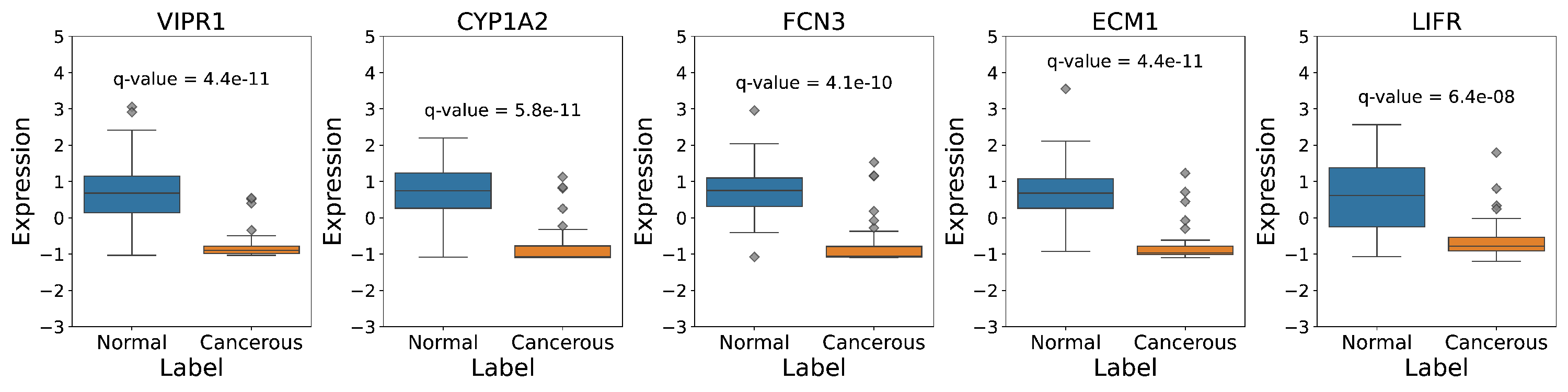}
\put(-430,115){\textbf{B}} 
\caption{}
\end{subfigure}
\hfill
\begin{subfigure}[b]{\textwidth}
\centering
\includegraphics[width=\textwidth]{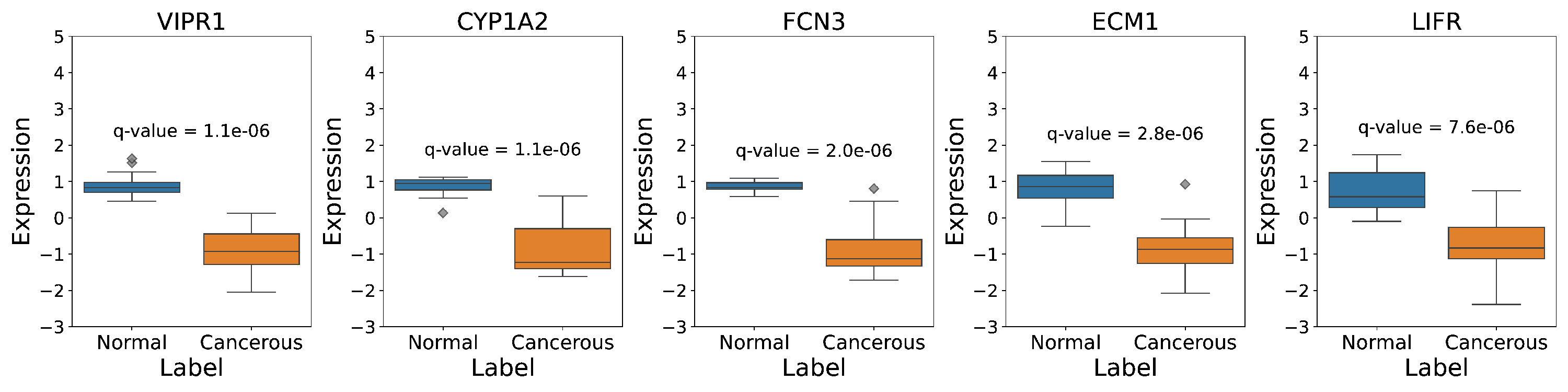}
\put(-430,115){\textbf{C}} 
\caption{}
\end{subfigure}
\caption{Box plots of gene expressions of V1PR1, CYP1A2, FCN3, ECM1 and LIFR in normal and cancerous tissues. (A) GSE25097, (B) GSE57957, (C) GSE60502.}
\label{fig:boxplots_genes}
\end{figure*}

\subsection{The Proposed Formula Outperforms State-of-the-Art Classifiers in HCC Detection}

To evaluate the effectiveness and generalizability capability of our method, we compared it with several state-of-the-art classifiers and an existing recent paper on HCC classification. The results highlight the advantage of our model compared to them.

First, we compared our model to four widely used classifiers: K-Nearest Neighbors (KNN) \cite{guo2003knn}, Random Forest (RF) \cite{breiman2001random}, Gradient Boosting (GB) \cite{friedman2001greedy}, and Support Vector Machines (SVM) \cite{hearst1998support}. Each classifier was trained on the GSE25097 dataset, which was also used to train our model, and then tested on multiple additional datasets: GSE60502, GSE57957, GSE64041, GSE121248, GSE47197, and GSE76297. For each method, we calculated the accuracy on each dataset. Moreover, we found the average performance of the accuracies over all datasets (see Table \ref{tab:average_performance}).

As it is shown, our method achieves the highest accuracy in three of the six datasets compared to the other classifiers. In the GSE60502, our method and KNN achieved similar accuracy, and KNN outperformed our method in the other two datasets. 

Although KNN outperformed our method in two datasets, our approach demonstrates consistent and stable results among all datasets, with accuracies exceeding 90\% in every case. In contrast, KNN shows significant variability in performance, with accuracies dropping as low as 77.42\%. Additionally, our method provides interpretability, a feature that traditional classifiers lack.  Furthermore, the results of traditional classifiers can vary depending on factors such as hyperparameter tuning, and other implementation details during training. In contrast, our method's predictions rely on a fixed formula that is independent of such parameters, ensuring reproducibility and robustness. Additionally, our method does not require machine learning expertise, as it uses a straightforward formula that can be directly applied to normalized gene expression values.

Our method demonstrates the highest average performance among all methods. This result underscores the robustness of our approach over diverse datasets, outperforming other classifiers in terms of generalizability and accuracy.

Next, we compared our model to an existing recent work \cite{zhang2024development} on HCC classification. This work which uses machine learning-based predictors for classifying HCC tissues from CwoHCC, has multiple drawbacks compared to our work.
\begin{itemize}
    \item Number of features: This work requires at least 18 secreted genes which may not be accessible in all datasets, whereas our model uses only five, making it easier to implement.
    \item Performance: Sensitivity results for the existing model on GSE121248 are reported as 0.93, 0.94, and 1 using mRMR+KNN, mRMR+SVM, and MRMD+SVM, respectively. On GSE47197, the sensitivities are 0.36, 0.98, and 1 for the same combinations. However, other critical metrics, such as specificity, accuracy, and F1 score, are not reported. These omissions make it difficult to evaluate the model's full performance and its ability to not misclassify normal samples as cancerous. In contrast, our model demonstrates a balanced performance, achieving a specificity of 0.97 on both GSE121248 and GSE47197 datasets while maintaining high sensitivities of 0.89 and 0.85, respectively. This balance ensures that our model not only identifies true positives effectively but also has a good capability of not misclassifying normal samples as cancerous.
\end{itemize}

\subsection{The Five Selected Genes Show Consistent Downregulation in HCC Tumor Samples}
The distribution of gene expression for  VIPR1, CYP1A2, FCN3, ECM1 and LIFR between normal and tumor samples is visualized using box plots (Figure \ref{fig:boxplots_genes} and Supplementary Figure S3). The plots show that tumor samples exhibit low expression values for all five genes in all datasets. In contrast, normal samples display higher expression levels.  To statistically assess these differences, we applied the Mann–Whitney U test to each gene, followed by Benjamini–Hochberg correction to adjust for multiple testing. The resulting q-values, all of which were below $10^{-6}$, are displayed above each plot and indicate statistically significant downregulation of the five genes in tumor samples. This consistent pattern supports their biological relevance as tumor suppressors and validates their selection for inclusion in the predictive formula. These results are consistent with previous research where each of 
VIPR1, CYP1A2, FCN3, ECM1 and LIFR individually has been identified as downregulated in HCC. For instance, VIPR1 downregulation has been associated with increased HCC proliferation and progression. The VIP/VIPR1 signaling pathway has been shown to upregulate the expression of arginine anabolic enzymes while suppressing CAD-mediated pyrimidine biosynthesis, suggesting a role in the metabolic regulation of HCC \cite{fu2022activation}.
CYP1A2 is typically silenced in HCC tumor tissues. Higher CYP1A2 expression has been correlated with favorable clinical features, including lower serum AFP levels, reduced vascular invasion, and improved tumor-free survival, highlighting its potential protective role \cite{yu2021cyp1a2,ma2023fcn3}.
FCN3 expression is significantly reduced in both HCC tissues and hepatoma cell lines at the mRNA and protein levels. Studies have shown that FCN3 functions as a tumor suppressor, inhibiting tumor development and progression \cite{ma2023fcn3}.
ECM1 is involved not only in regulating malignant cell proliferation but also in enhancing cancer cell migration and invasion, contributing to tumor aggressiveness \cite{chen2016ecm1}.
LIFR has been identified as a metastasis suppressor in HCC. It exerts its effects by negatively regulating the PI3K–AKT–MMP13 signaling axis, thereby limiting metastatic potential \cite{luo2015lifr}. 
\subsection{Differential Gene Expression Analysis and Correlations}
A volcano plot (Supplementary Figure S4 top row) visualizes the differential expression of genes based on log2 fold change and -log10 p-value. The five highlighted genes—VIPR1, CYP1A2, FCN3, ECM1, and LIFR—are marked with bold labels and a purple color. The plot includes dashed lines to represent thresholds for statistical significance, distinguishing significant genes from non-significant ones. The behavior of the highlighted genes varies across datasets. In GSE60502, all five genes are significantly downregulated. However, in GSE57957, ECM1 and LIFR do not reach statistical significance. This highlights a limitation of traditional techniques, which may overlook the importance of some genes. However, our results indicates the relevance of these genes and demonstrates their strong predictive capability for HCC classification (for more results see Supplementary Figure S4 top row).

The heatmap (Supplementary Figure S4 bottom row) shows the correlation matrix of gene expression levels across different datasets. Using a cool-to-warm color palette, the heatmap highlights high correlation among the five genes on multiple datasets. On the GSE47197, FCN3 shows low correlation with other genes. However, other genes still have high correlations. This suggests a potential shared regulatory mechanism or biological pathway in tumor suppression.

\subsection{Several Genes in the Predictive Formula Are Linked to Known Drugs}
Table S1 presents the drugs associated with the five genes used in our predictive formula, as derived from DrugBank \cite{wishart2018drugbank}. VIPR1 is linked to one drug, while CYP1A2 is associated with over 300 drugs, of which a few representative examples are listed in the table. FCN3 has been associated with two drugs. However, ECM1 and LIFR are not currently recognized as potential drug targets, suggesting opportunities for further investigation into their therapeutic potential. These findings provide valuable insights into the relevance of these genes in therapeutic contexts and highlight the need for additional research, particularly for ECM1 and LIFR.
\section{Conclusion}
In this study, we introduced a novel predictive formula for HCC identification based on the expression levels of five key genes: VIPR1, CYP1A2, FCN3, ECM1, and LIFR. We derived this formula by training the KAN model on GSE25097. Not only the presented formula is simple, accessible and interpretable, it achieves high accuracy on seven independent datasets, highlighting its generalizability and potential for clinical application. Additionally, the findings emphasize the combined significance of these five genes as biomarkers for HCC detection, presenting new insights into their roles in tumor suppression. 

Despite the performance of this formula in classifying normal from HCC cancerous tissues, the study has some limitations. While computational and statistical results are robust, there is a need for mechanistic explorartion of the biological roles and the interrelationships of the five genes as tumor suppressors. 

\section*{Data Availability}
All data are publicly available and cited in the paper.
\section*{Conflict of Interest}
None declared.
\section*{CRediT Authorship Contribution Statement}
\textbf{Aram Ansary Ogholbake:} Conceptualization, Formal Analysis, Methodology, Software, Visualization, Writing – original draft, Writing – review \& editing
\textbf{Qiang Cheng:} Conceptualization, Funding acquisition, Methodology, Supervision, Writing – review \& editing
\section*{Acknowledgments}
This research is supported by the National Artificial Intelligence Research Resource (NAIRR) Pilot NSF OAC 240219, and Jetstream2, Bridges2, and Neocortex Resources. Additional support was provided by the NSF  [IIS 2327113, ITE 2433190] and the NIH [R21AG070909,
P30AG072946, R01HD101508-01].

\bibliographystyle{plain} 
\bibliography{cas-refs}

\renewcommand{\thefigure}{S\arabic{figure}}
\setcounter{figure}{0}
\renewcommand{\tablename}{Table}
\renewcommand{\thetable}{S\arabic{table}}
\setcounter{table}{0}

\begin{figure*}[]
\centering
\begin{subfigure}[b]{0.3\textwidth}
\centering
\includegraphics[width=\textwidth]{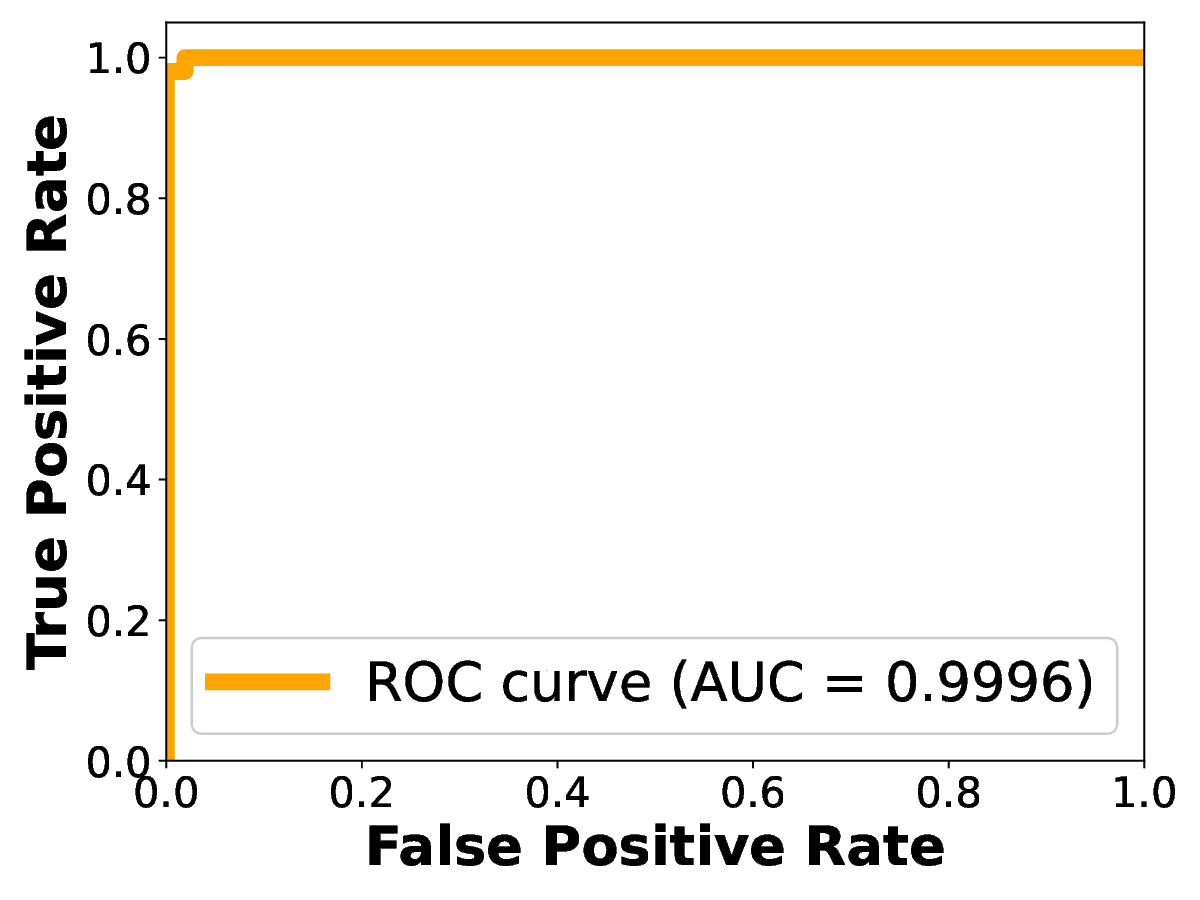}
\put(-120,115){\textbf{A}} 
\end{subfigure}
\hfill
\begin{subfigure}[b]{0.3\textwidth}
\centering
\includegraphics[width=\textwidth]{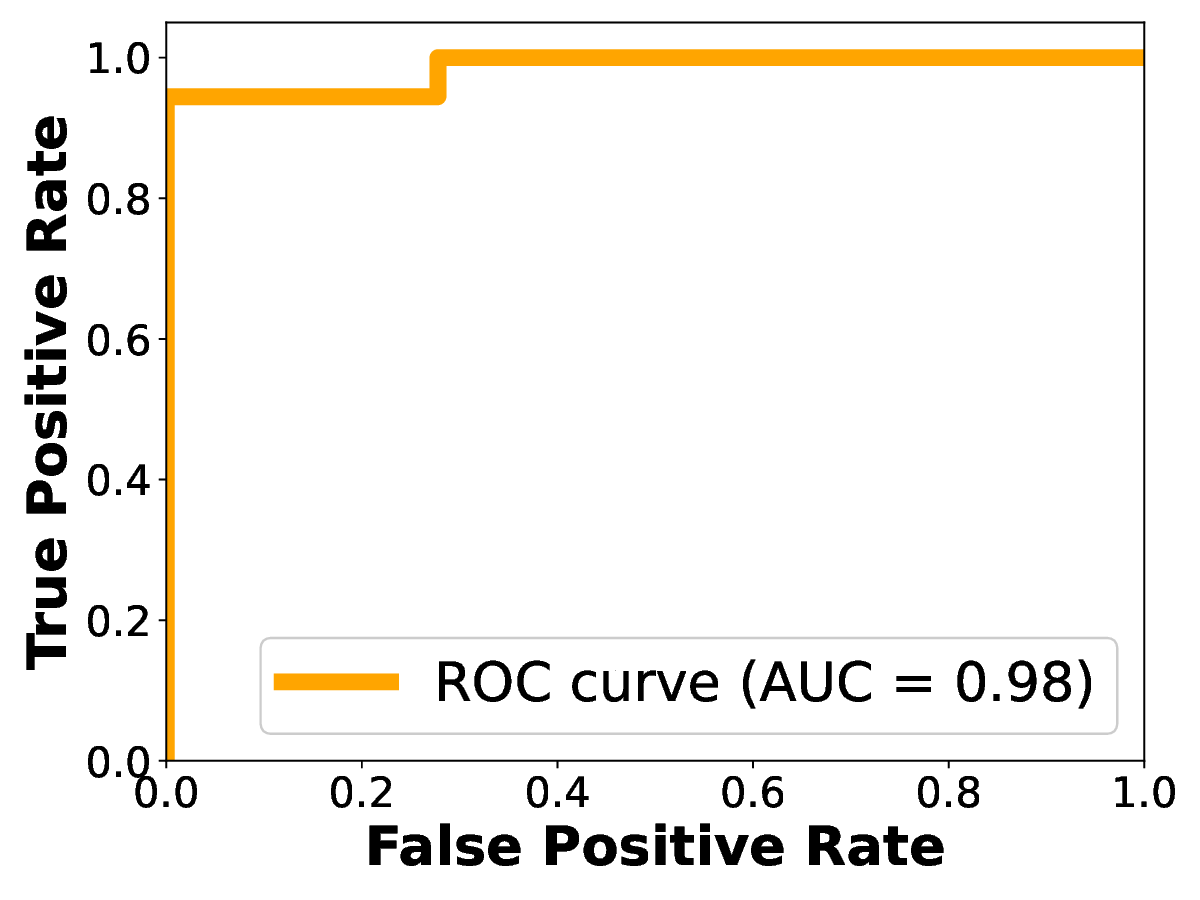}
\put(-120,115){\textbf{B}} 
\end{subfigure}
\hfill
\begin{subfigure}[b]{0.3\textwidth}
\centering
\includegraphics[width=\textwidth]{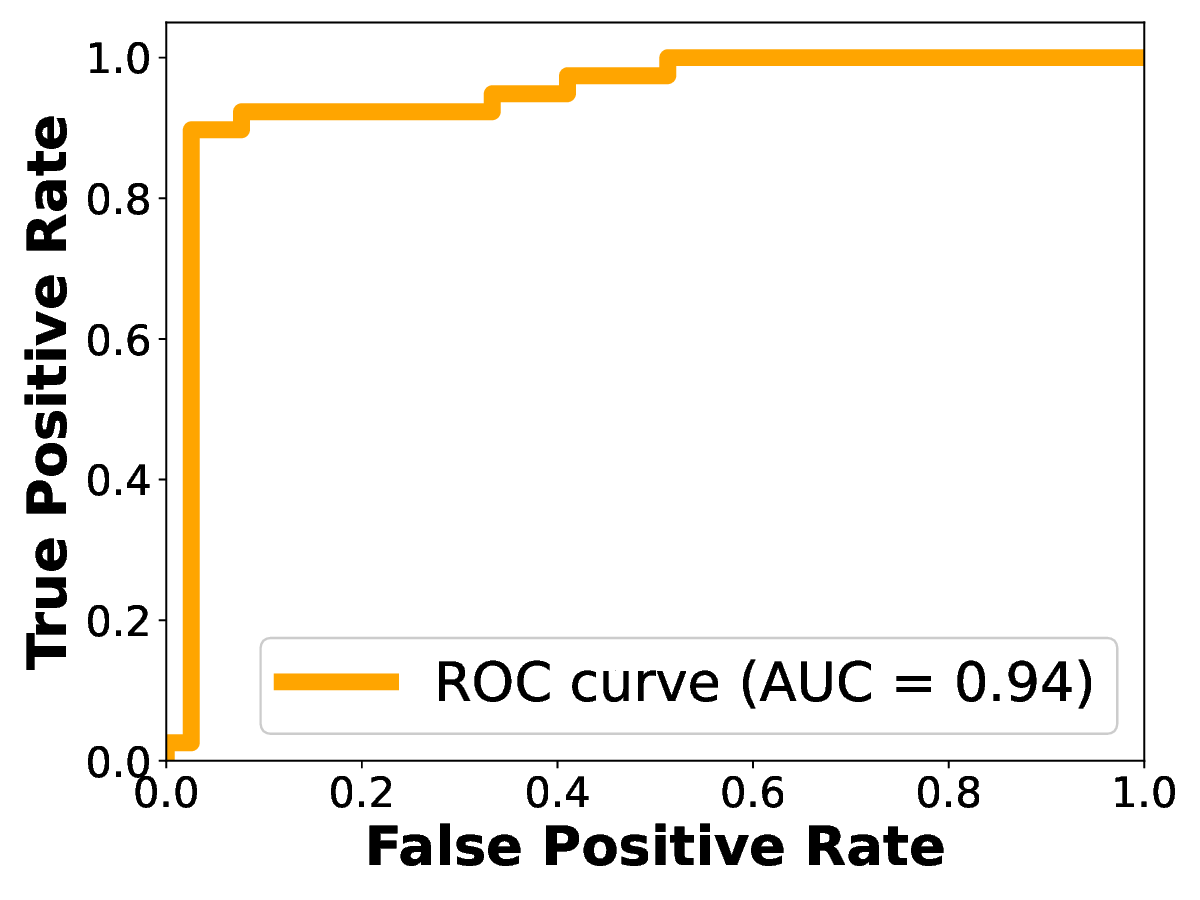}
\put(-120,115){\textbf{C}} 
\end{subfigure}
\vspace{1em} 
 \begin{subfigure}[b]{0.3\textwidth}
\centering
\includegraphics[width=\textwidth]{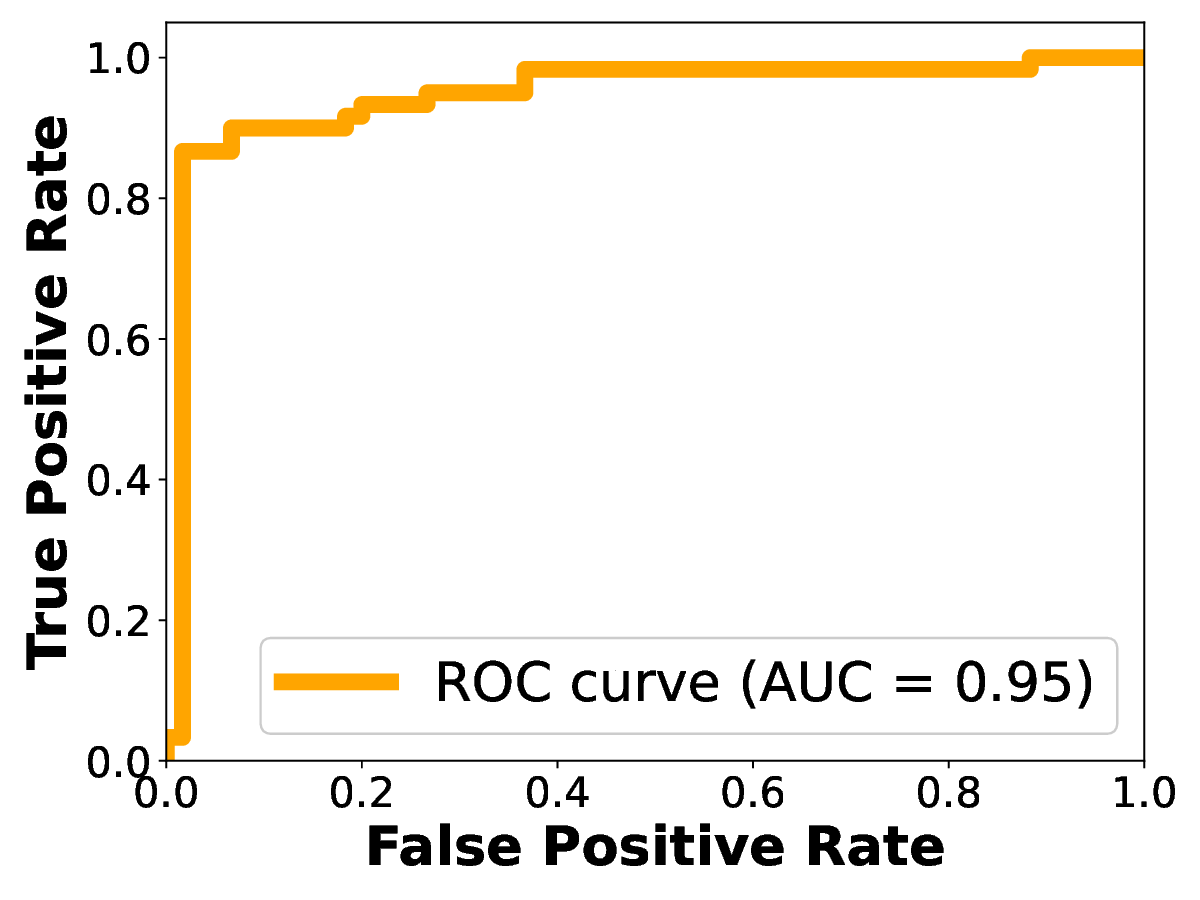}
\put(-120,115){\textbf{D}} 
\end{subfigure}
\begin{subfigure}[b]{0.3\textwidth}
 \centering
 \includegraphics[width=\textwidth]{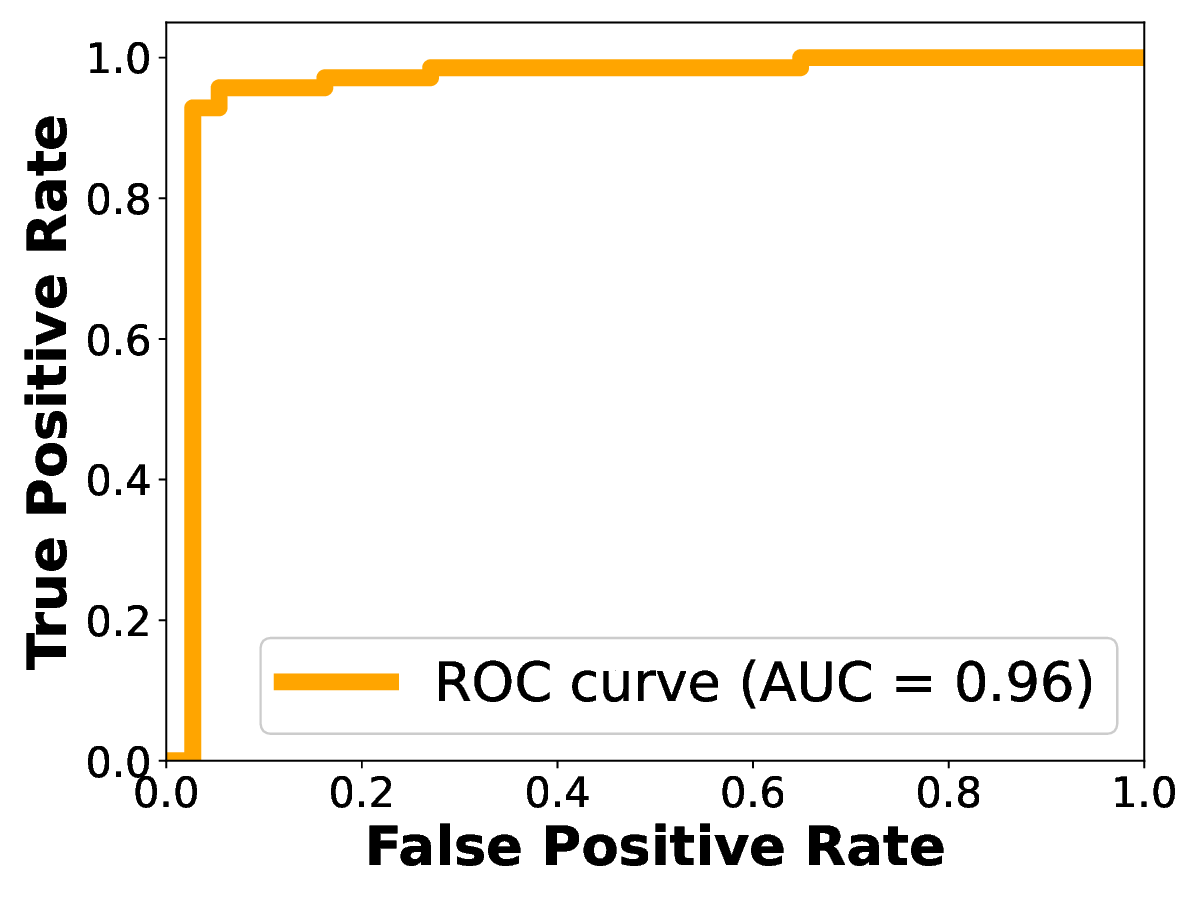}
 \put(-120,115){\textbf{E}} 
\end{subfigure}
\begin{subfigure}[b]{0.3\textwidth}
\centering
\includegraphics[width=\textwidth]{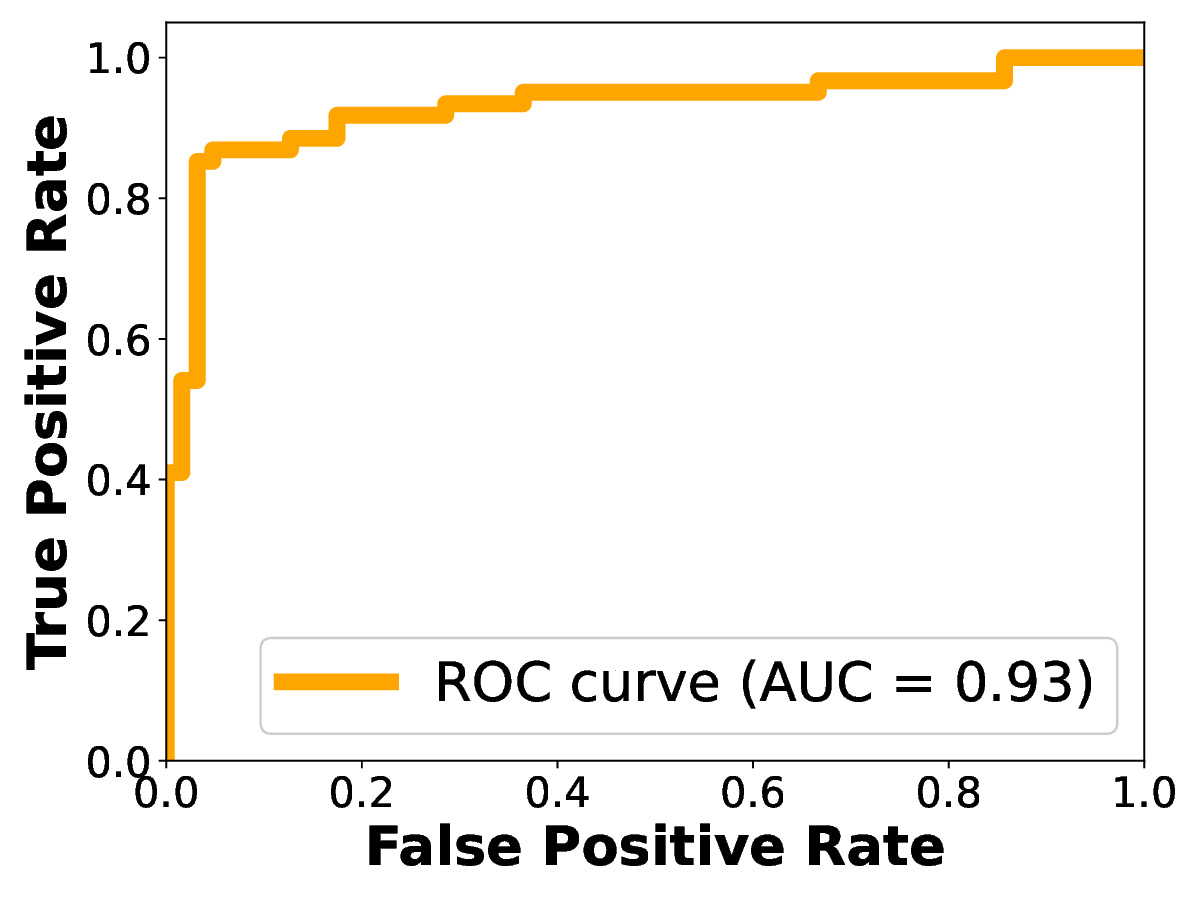}
\put(-120,115){\textbf{F}} 
\end{subfigure}
\vspace{1em}
\begin{subfigure}[b]{0.3\textwidth}
\centering
\includegraphics[width=\textwidth]{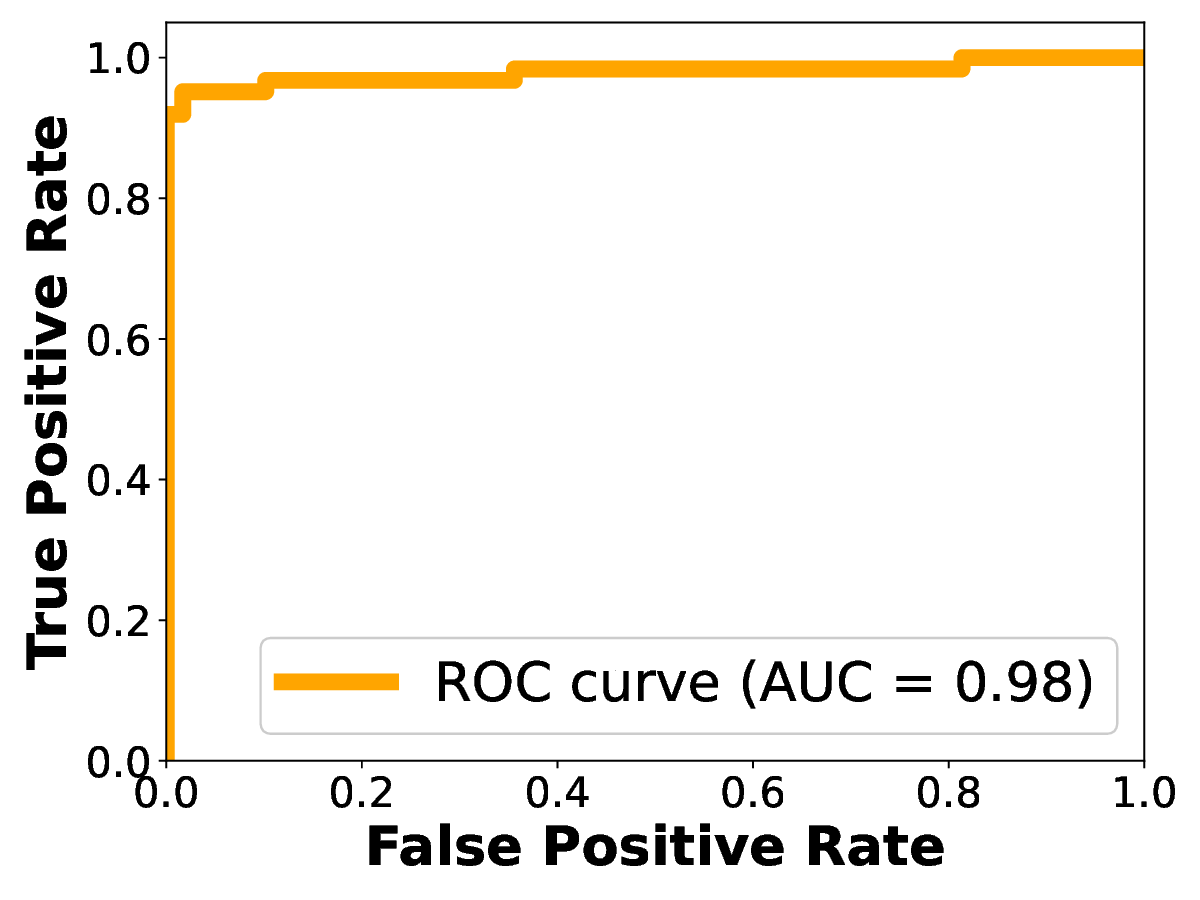}
\put(-120,115){\textbf{G}} 
\end{subfigure}
\caption{ROC curves for different datasets. (A) Validation result on the training data (GSE25097); (B)-(G) Test results on independent datasets of GSE60502, GSE57957, GSE64041, GSE121248, GSE47197, and GSE76297.}
\label{fig:roc}
\end{figure*}

\begin{figure*}[h!]
    \centering

    \begin{subfigure}{0.22\textwidth}
        \centering
        \includegraphics[width=\textwidth]{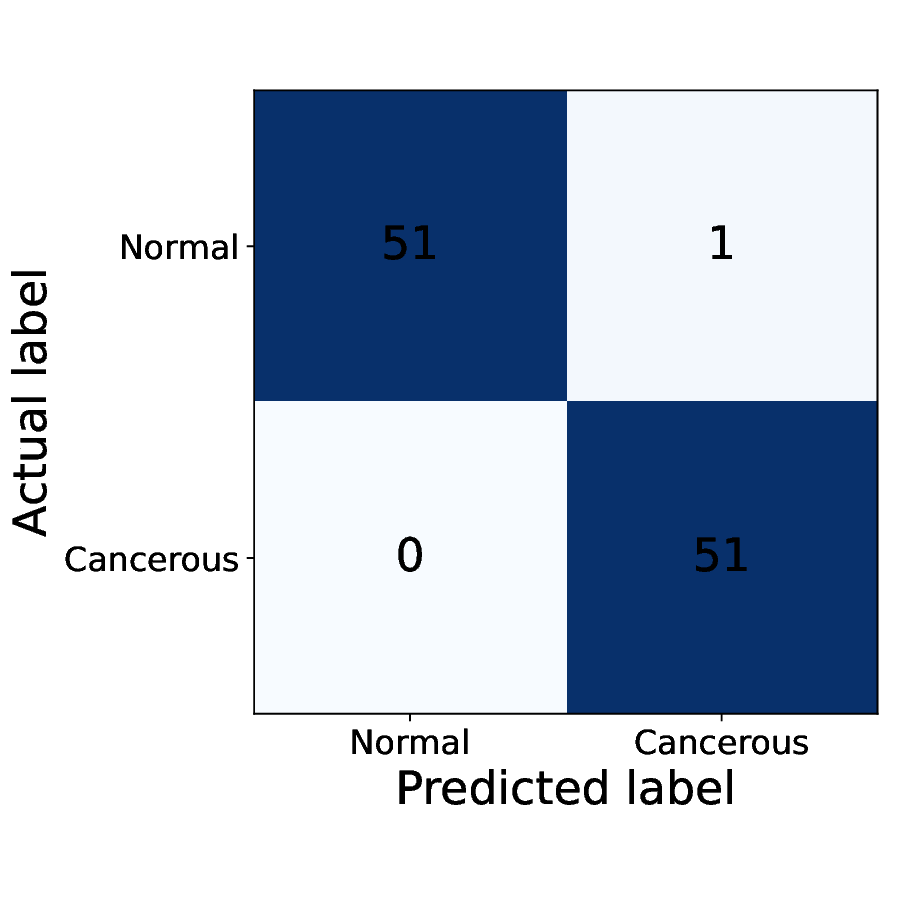}
    \put(-100,100){\textbf{A}} 
        \label{fig:confusion1}
    \end{subfigure}
    \hfill
    \begin{subfigure}{0.22\textwidth}
        \centering
        \includegraphics[width=\textwidth]{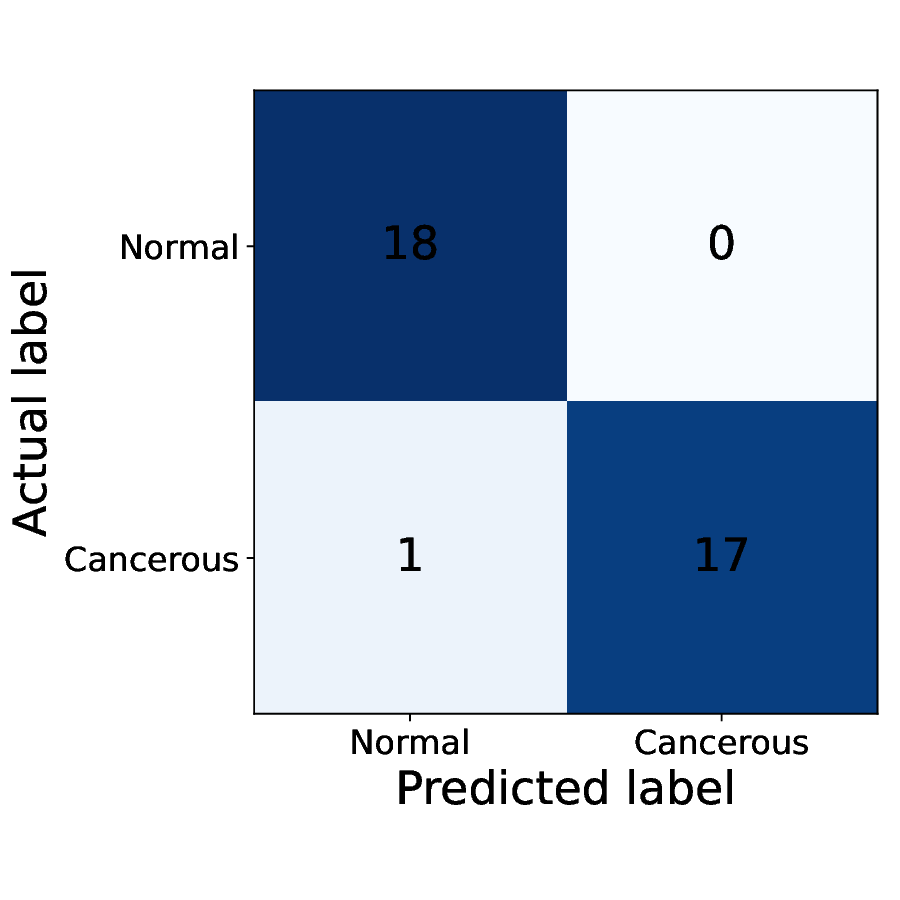}
    \put(-100,100){\textbf{B}} 
    \end{subfigure}
    \hfill
    \begin{subfigure}{0.22\textwidth}
        \centering
        \includegraphics[width=\textwidth]{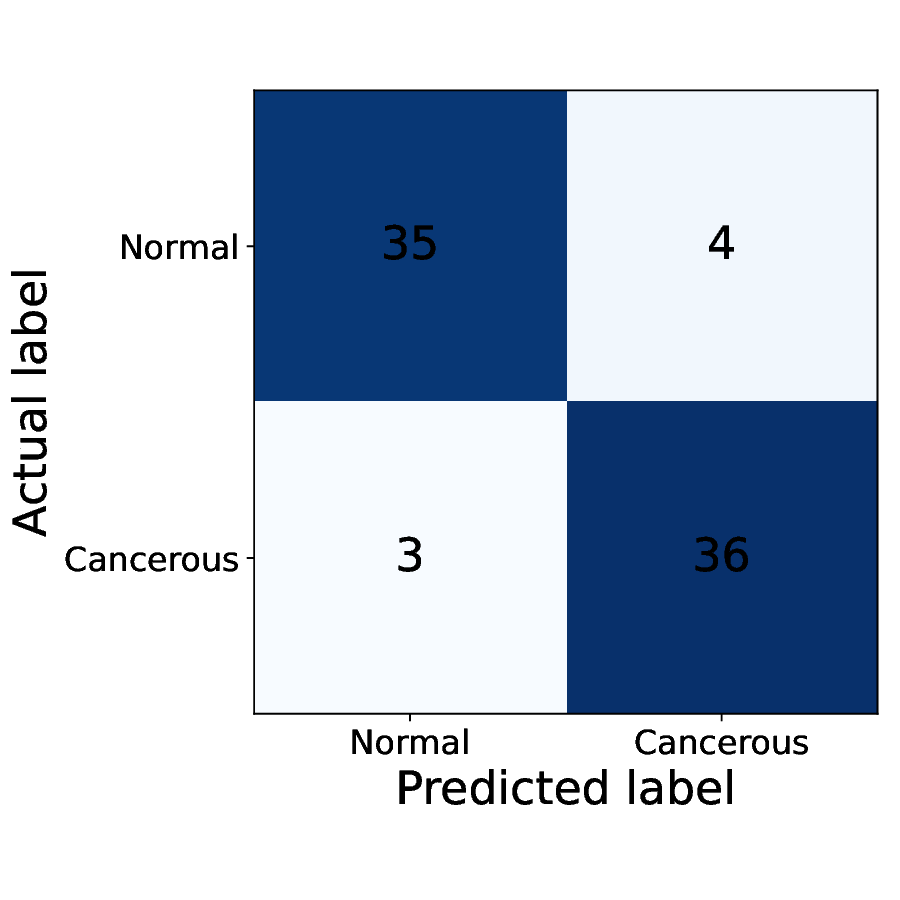}
    \put(-100,100){\textbf{C}} 
    \end{subfigure}
        \begin{subfigure}{0.22\textwidth}
        \centering
        \includegraphics[width=\textwidth]{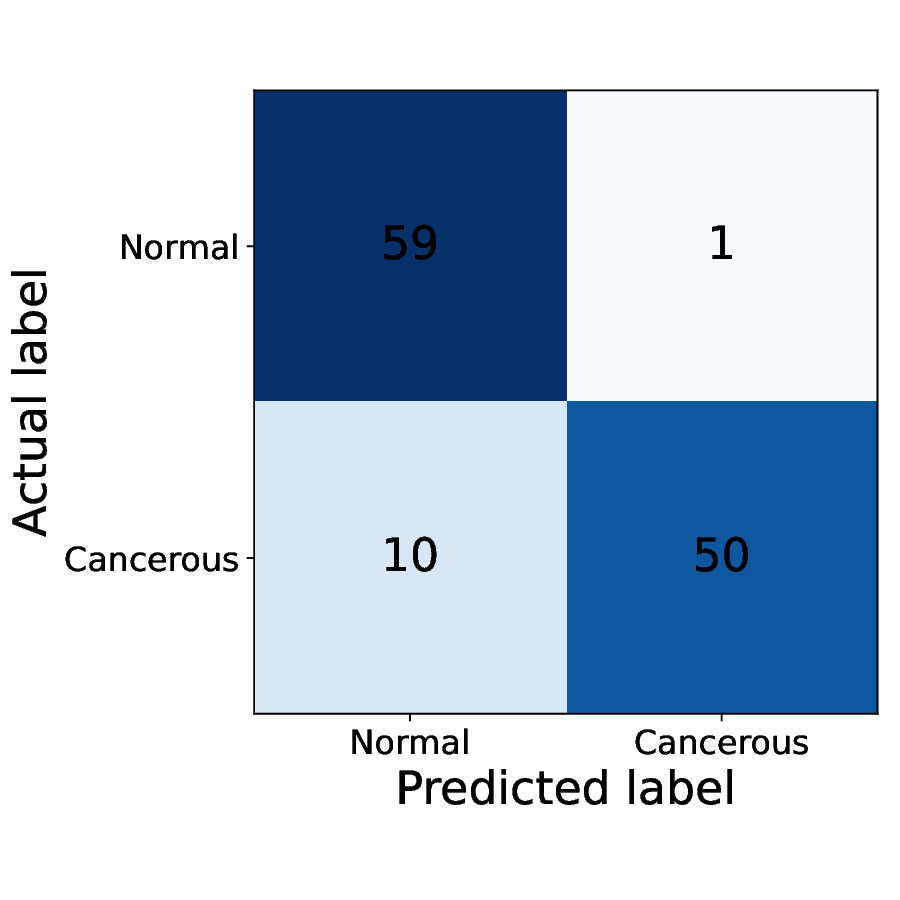}
    \put(-100,100){\textbf{D}} 
    \end{subfigure}
        \hfill

    \begin{subfigure}{0.22\textwidth}
        \centering
        \includegraphics[width=\textwidth]{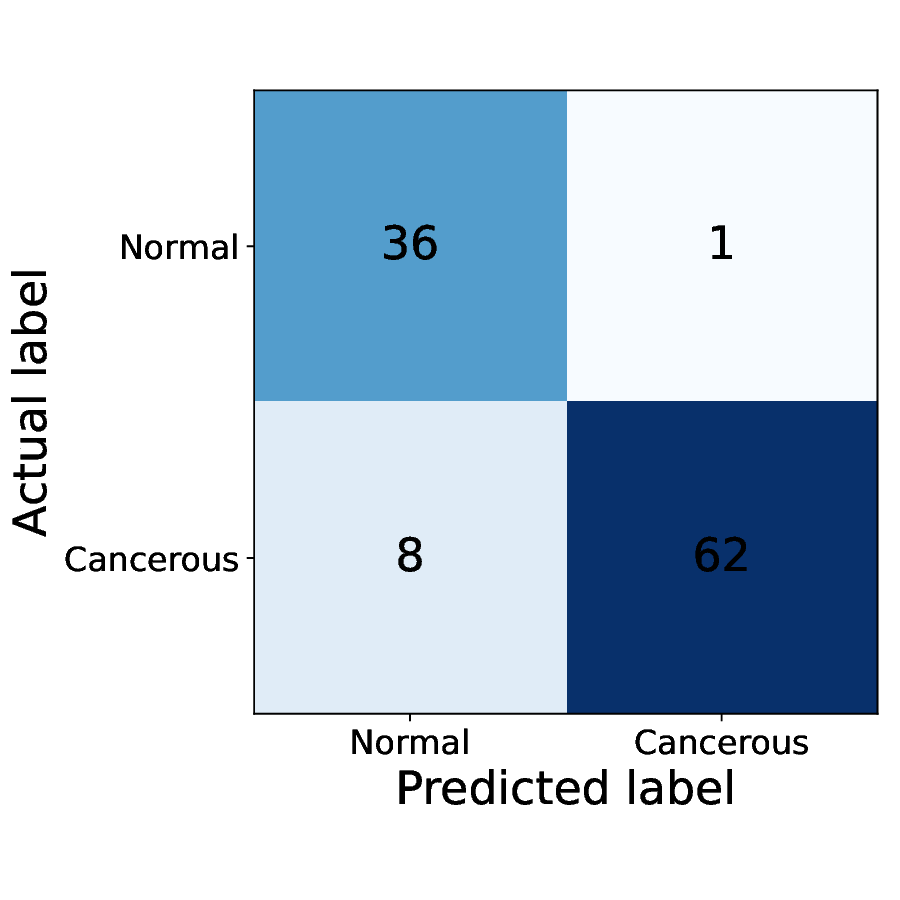}
    \put(-90,100){\textbf{E}} 
    \end{subfigure}
        \begin{subfigure}{0.22\textwidth}
        \centering
        \includegraphics[width=\textwidth]{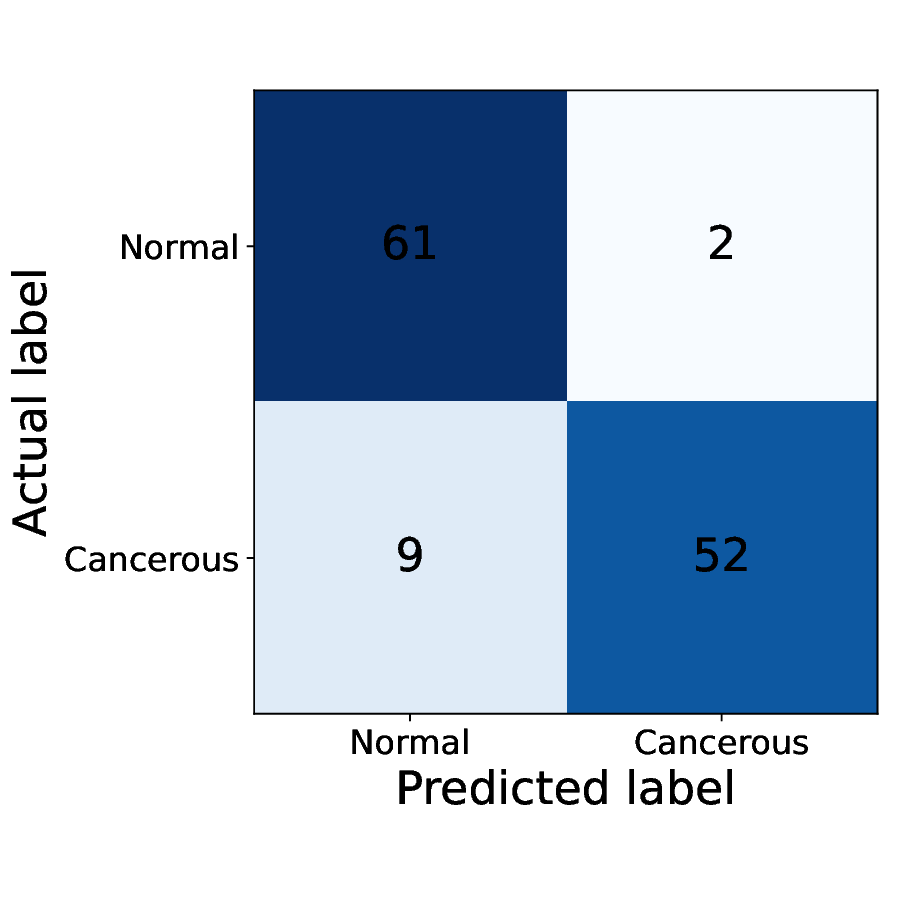}
    \put(-90,100){\textbf{F}} 
    \end{subfigure}
        \begin{subfigure}{0.22\textwidth}
        \centering
        \includegraphics[width=\textwidth]{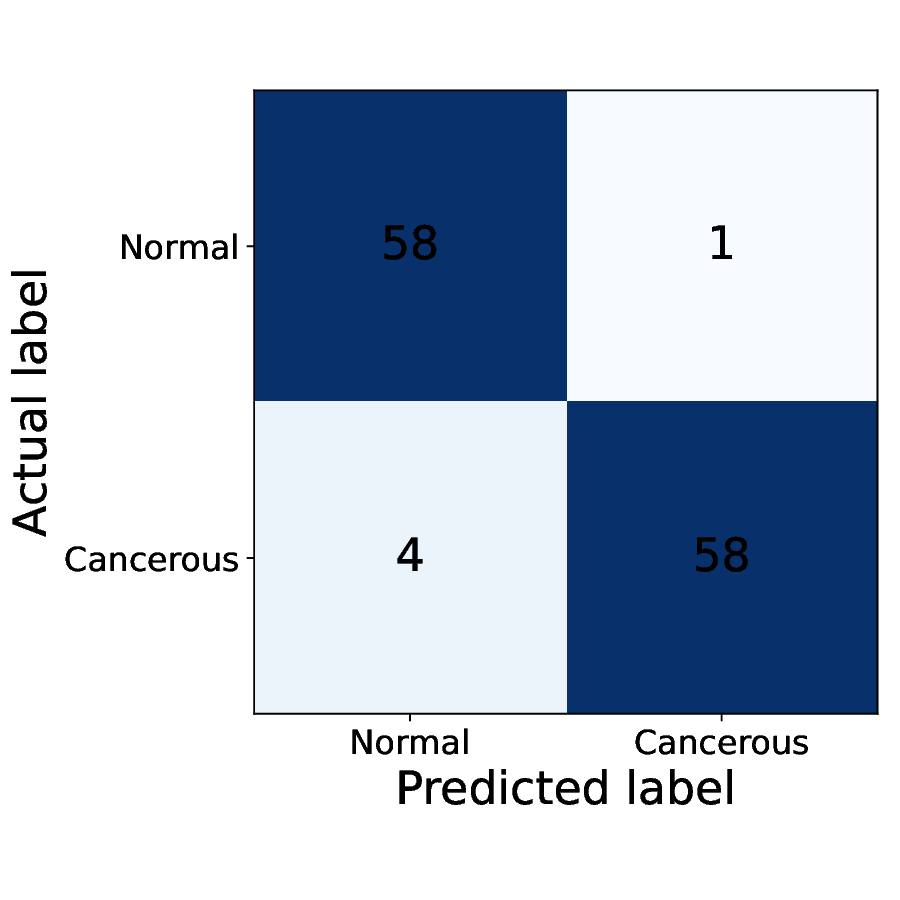}
    \put(-90,100){\textbf{G}} 
    \end{subfigure}
    \caption{Confusion matrices for different datasets on (A) GSE25097, (B) GSE60502, (C) GSE57957, (D) GSE64041 (E) GSE121248, (F) GSE47197, and (G) GSE76297.}
\end{figure*}

\begin{figure*}[]
    \centering
    \begin{subfigure}[b]{\textwidth}
        \centering
        \includegraphics[width=\textwidth]{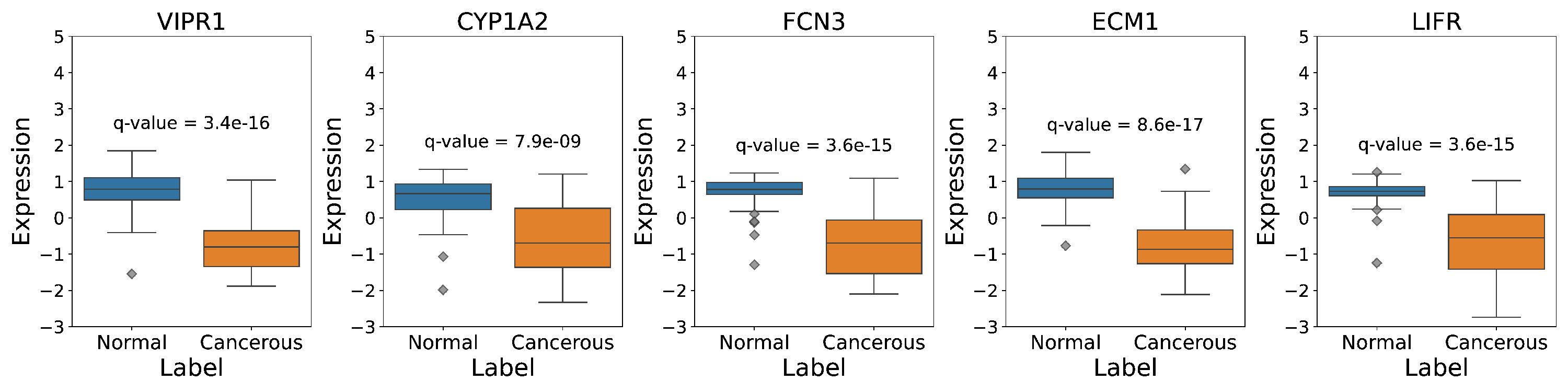}
    \put(-430,115){\textbf{A}} 
    \end{subfigure}
    \hfill
        \begin{subfigure}[b]{\textwidth}
        \centering
        \includegraphics[width=\textwidth]{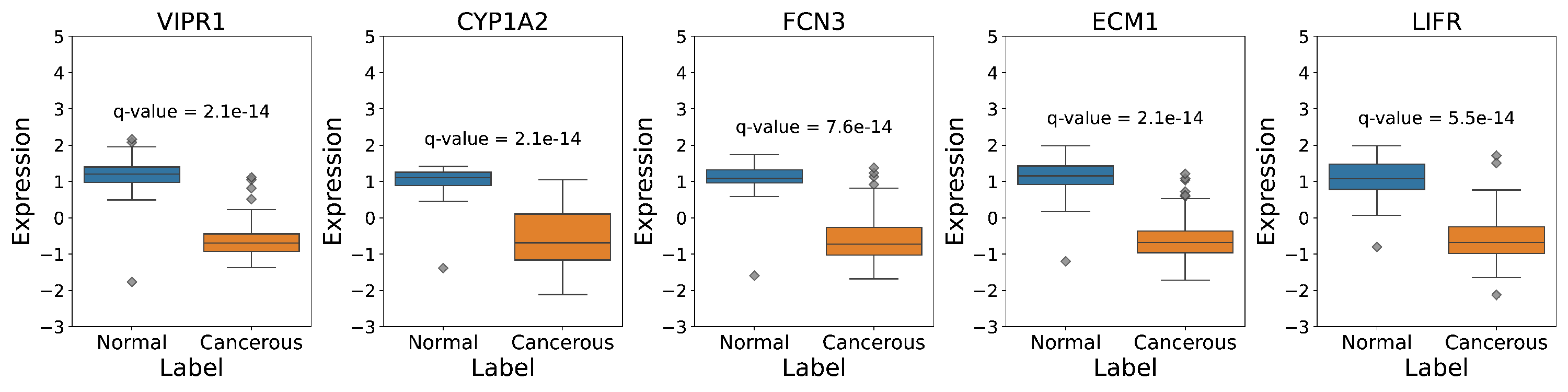}
    \put(-430,115){\textbf{B}} 
    \end{subfigure}
    \hfill
    \begin{subfigure}[b]{\textwidth}
        \centering
        \includegraphics[width=\textwidth]{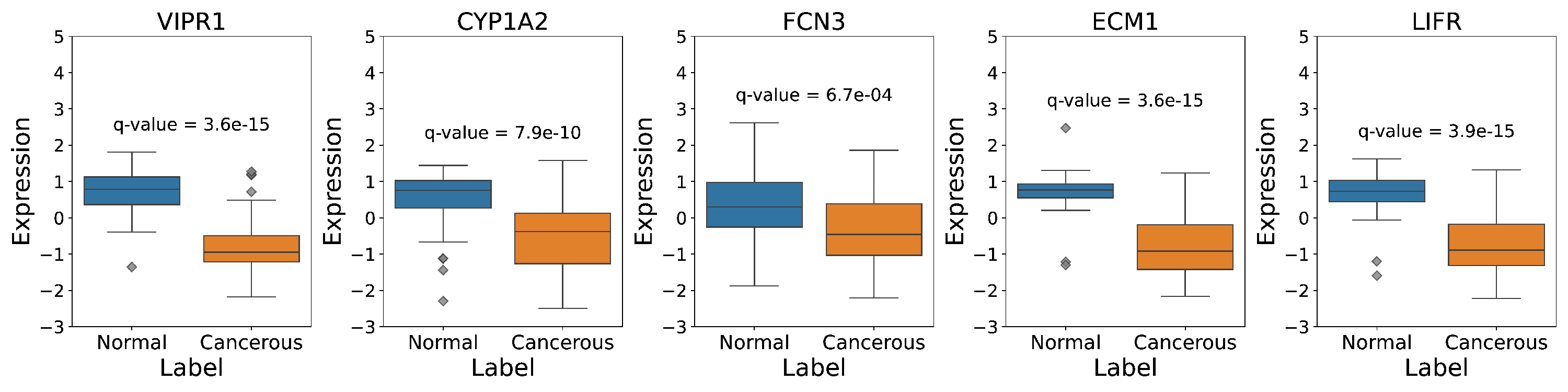}
    \put(-430,115){\textbf{C}} 
    \end{subfigure}
    \hfill
        \begin{subfigure}[b]{\textwidth}
        \centering
        \includegraphics[width=\textwidth]{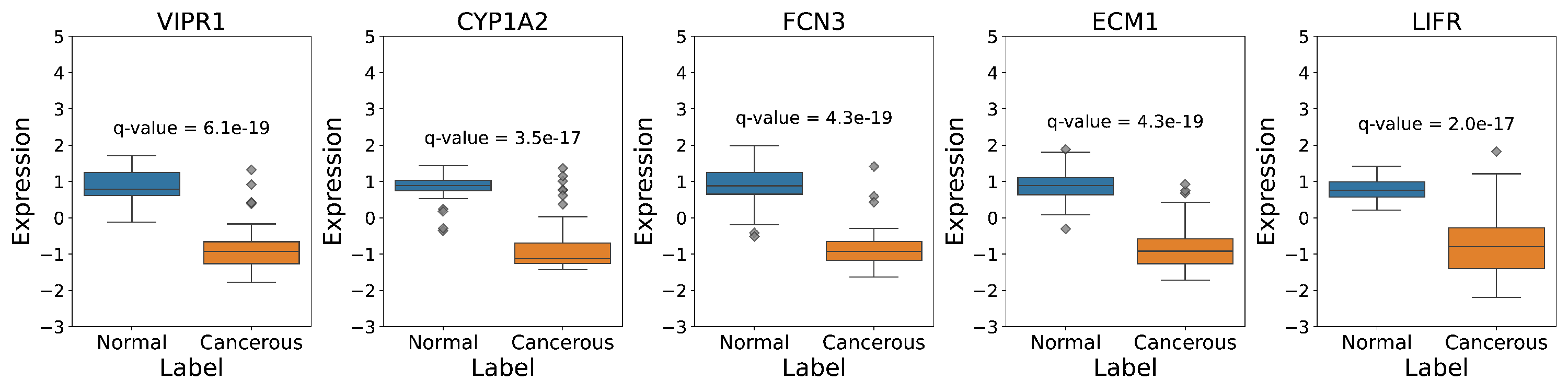}
    \put(-430,115){\textbf{D}} 
    \end{subfigure}
    \caption{Box plots of gene expressions of V1PR1, CYP1A2, FCN3, ECM1 and LIFR in normal and cancerous tissues. (A) GSE64041, (B) GSE121248, (C) GSE47197, and (D) GSE76297.}
\end{figure*}

\begin{figure*}[]
\centering

\begin{subfigure}[b]{0.3\textwidth}
    \setlength{\unitlength}{1pt}
    \begin{picture}(0,0)
        \put(-10,10){\textbf{A}}
    \end{picture}
    \includegraphics[width=\textwidth]{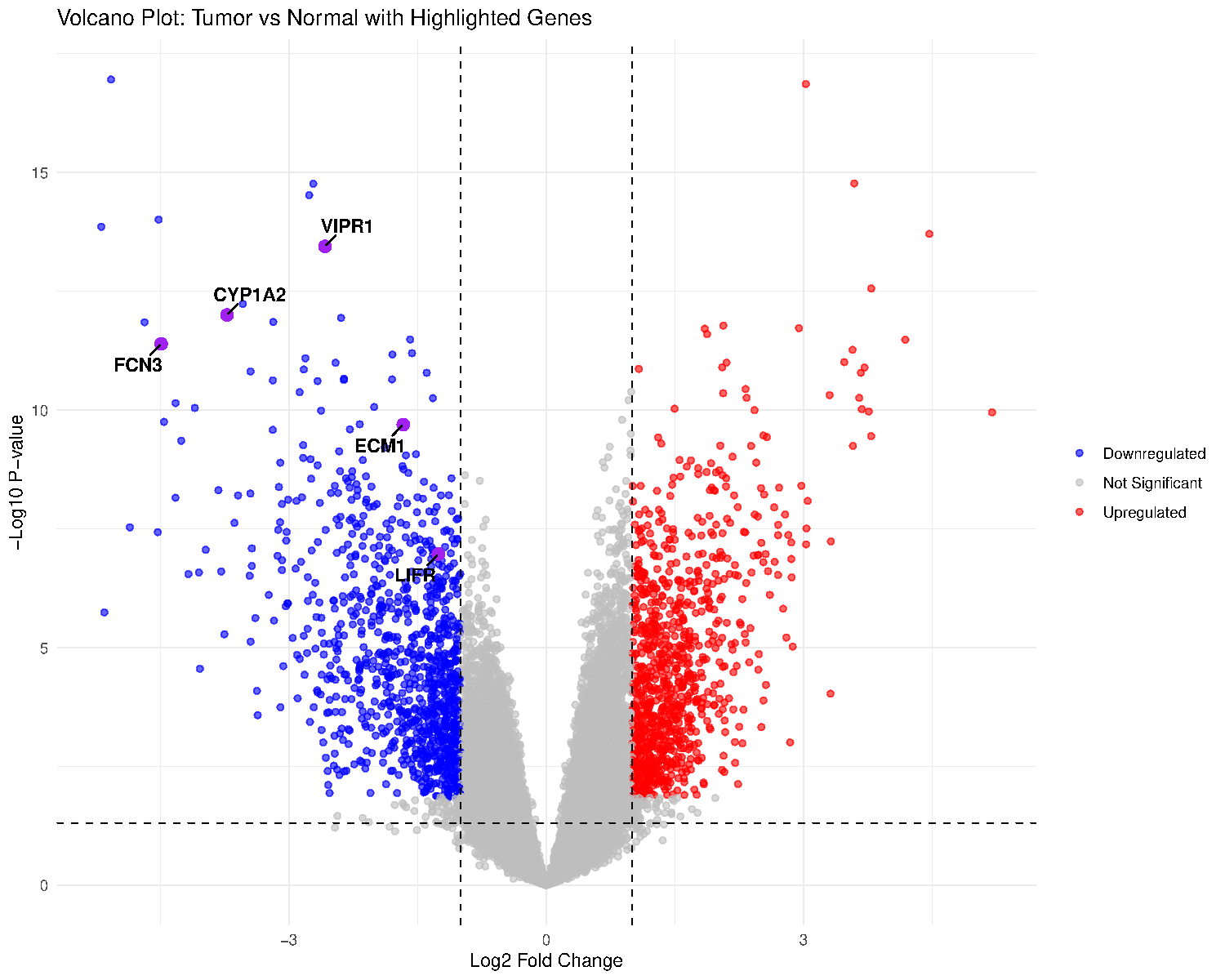}
    \includegraphics[width=\textwidth]{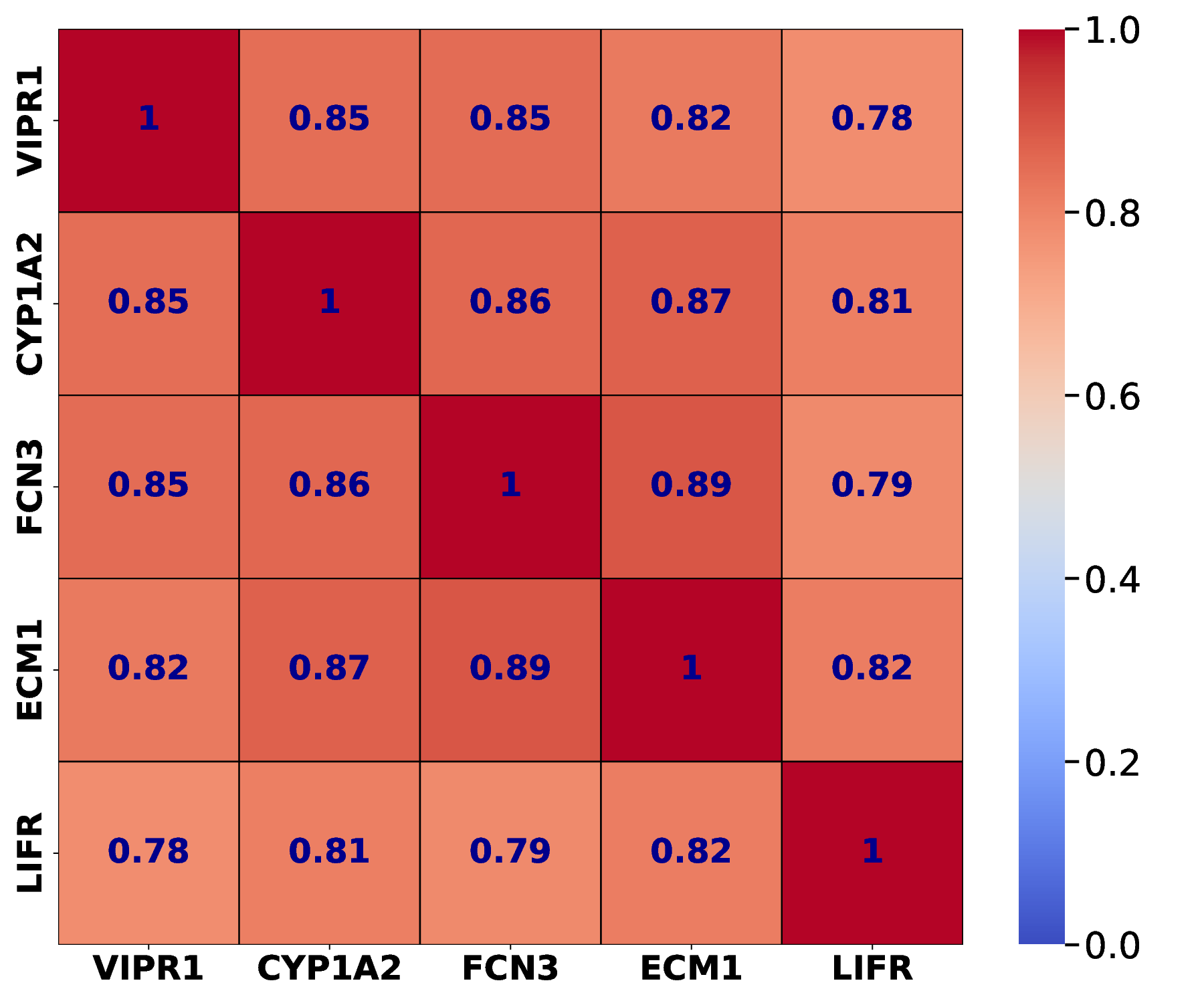}
\end{subfigure}
\hfill
\begin{subfigure}[b]{0.3\textwidth}
    \setlength{\unitlength}{1pt}
    \begin{picture}(0,0)
        \put(-10,10){\textbf{B}}
    \end{picture}
    \includegraphics[width=\textwidth]{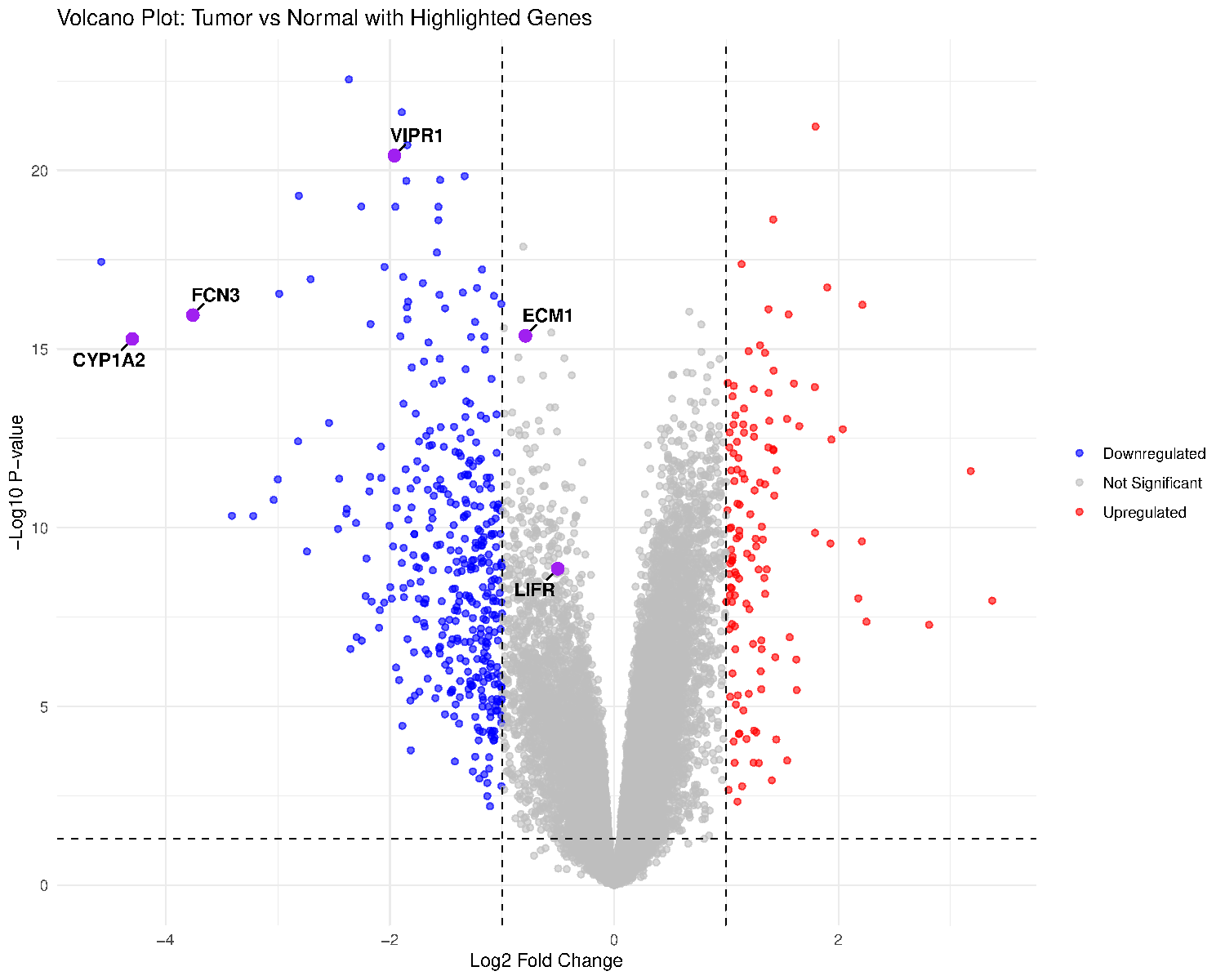}
    \includegraphics[width=\textwidth]{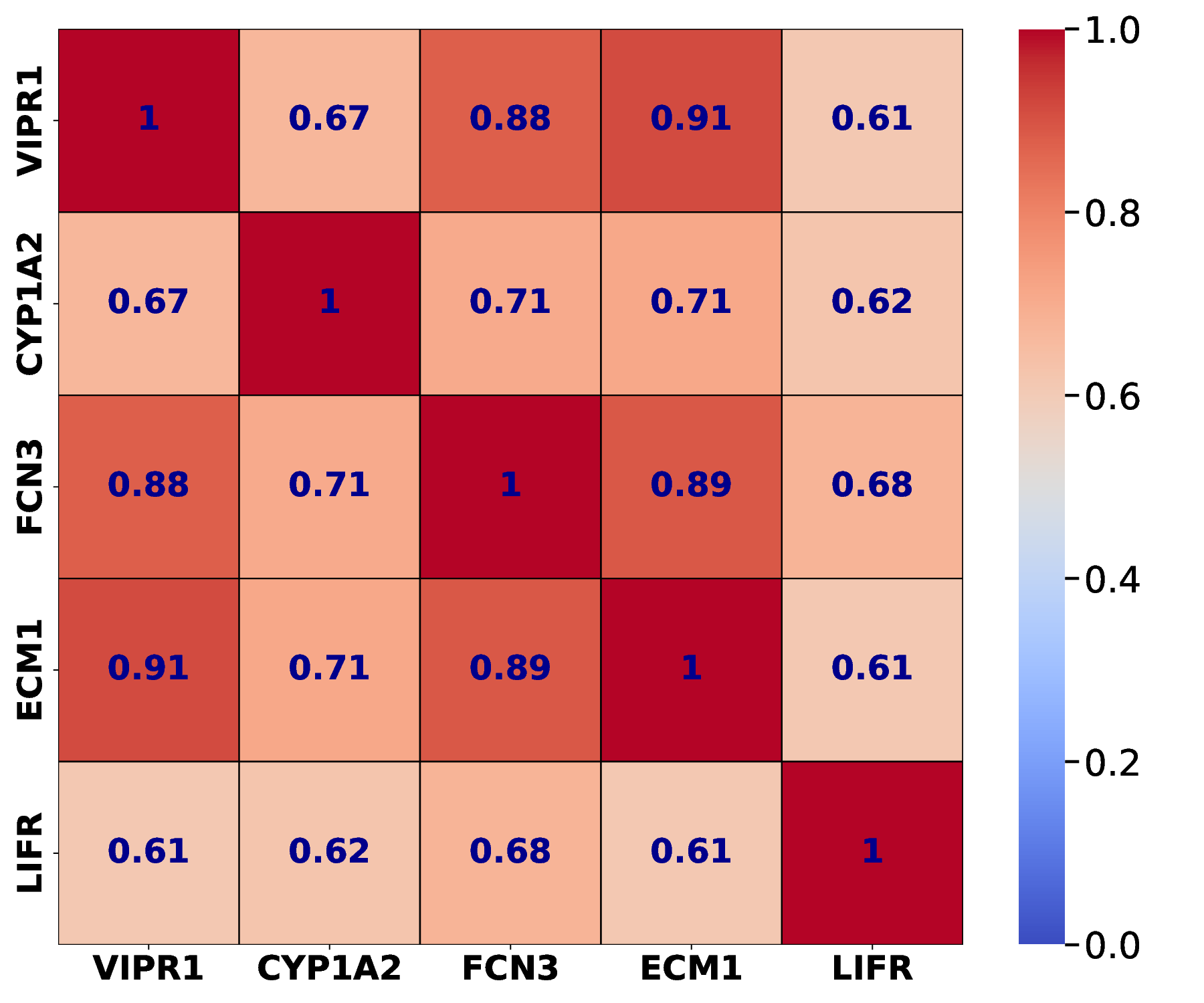}
\end{subfigure}
\hfill
\begin{subfigure}[b]{0.3\textwidth}
    \setlength{\unitlength}{1pt}
    \begin{picture}(0,0)
        \put(-10,10){\textbf{C}}
    \end{picture}
    \includegraphics[width=\textwidth]{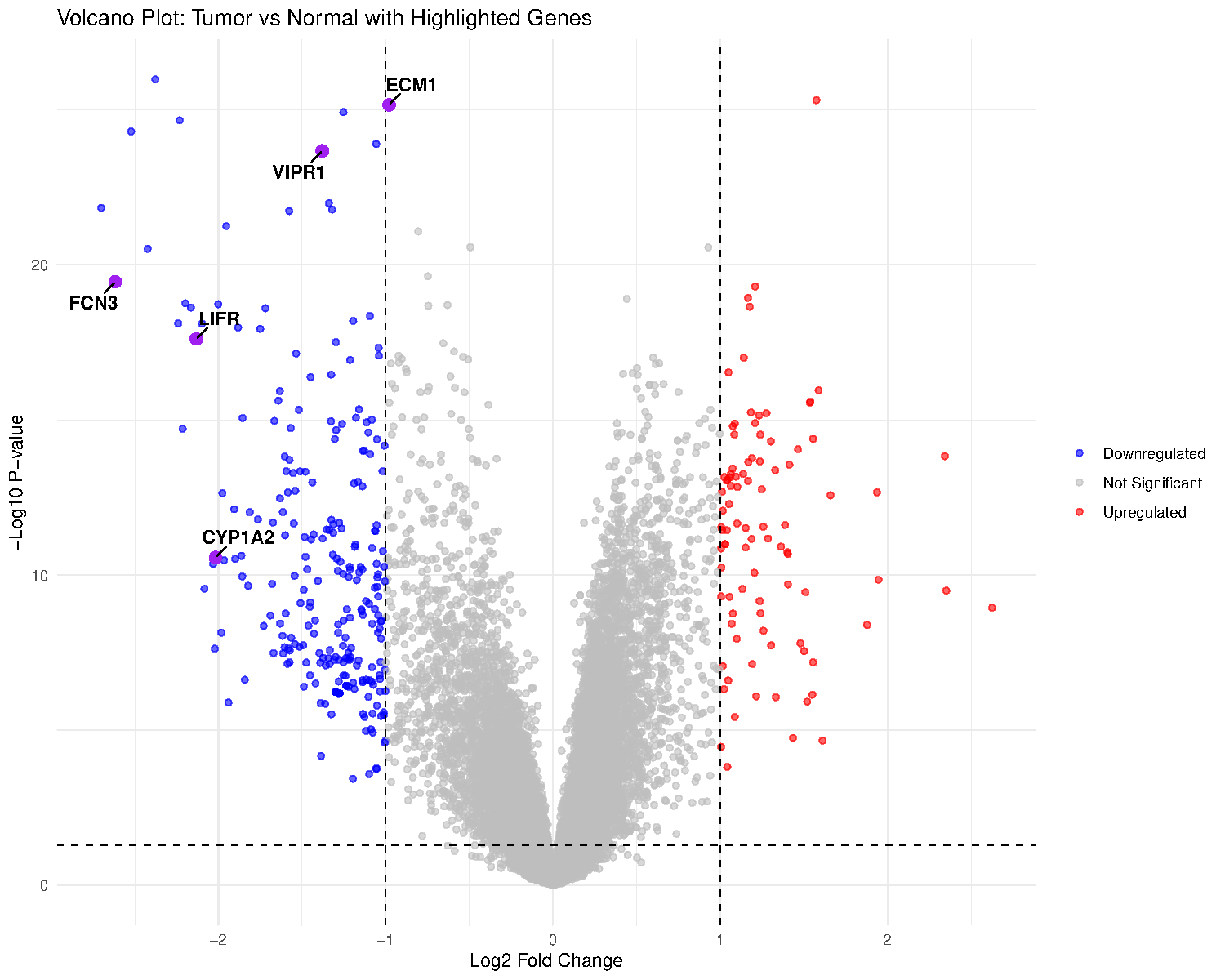}
    \includegraphics[width=\textwidth]{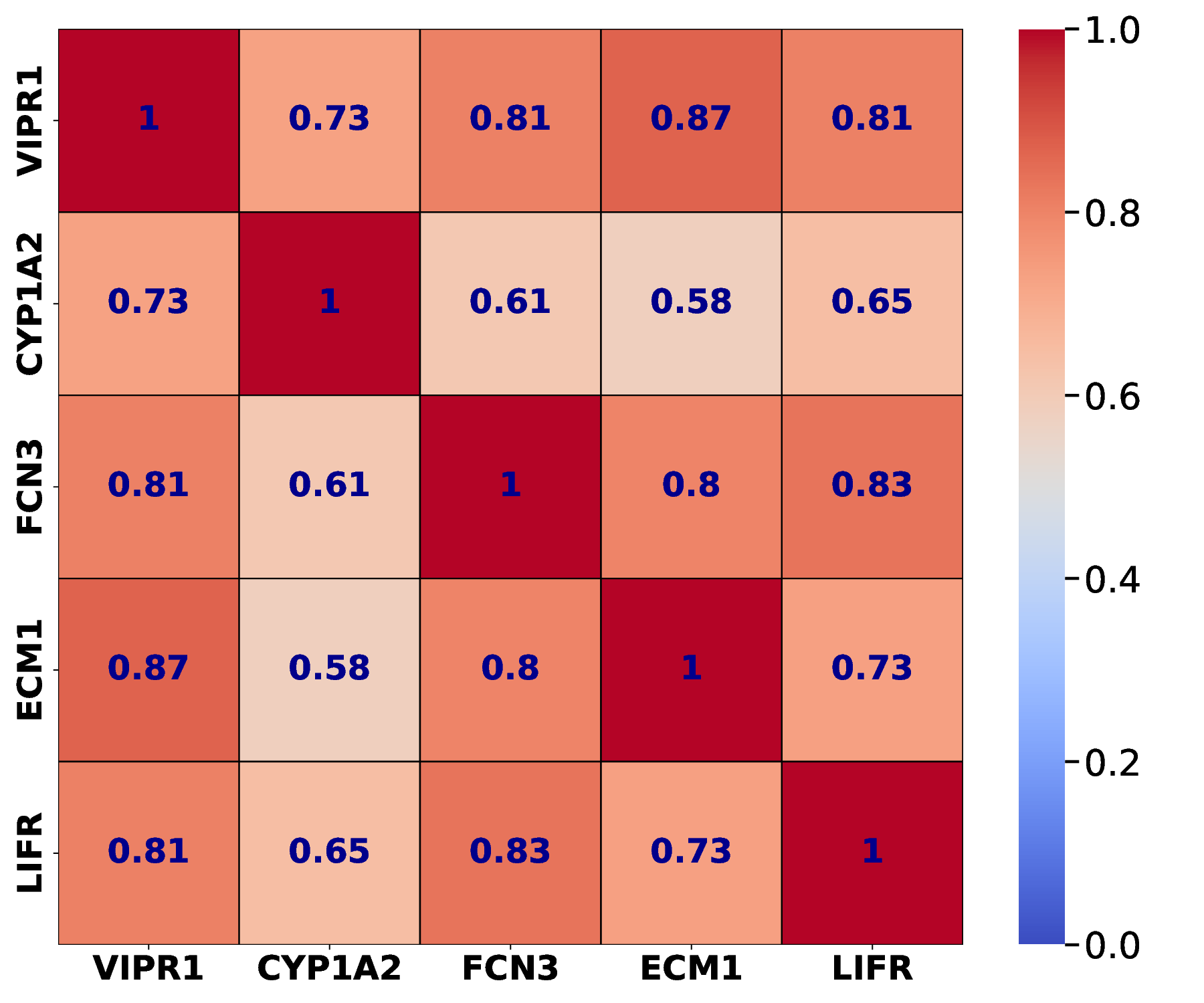}
\end{subfigure}

\vspace{1em}

\begin{subfigure}[b]{0.3\textwidth}
    \setlength{\unitlength}{1pt}
    \begin{picture}(0,0)
        \put(-10,5){\textbf{D}}
    \end{picture}
    \includegraphics[width=\textwidth]{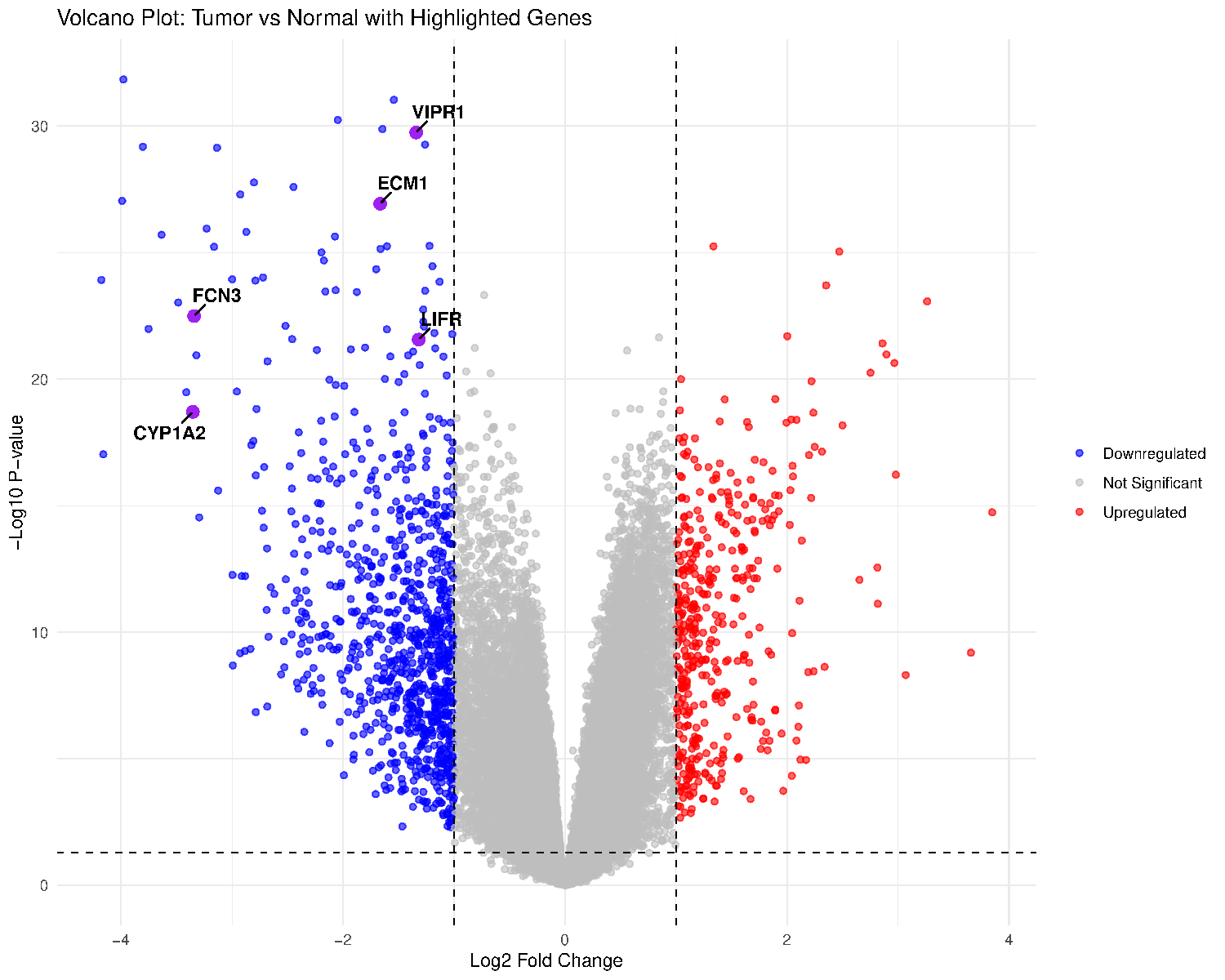}
    \includegraphics[width=\textwidth]{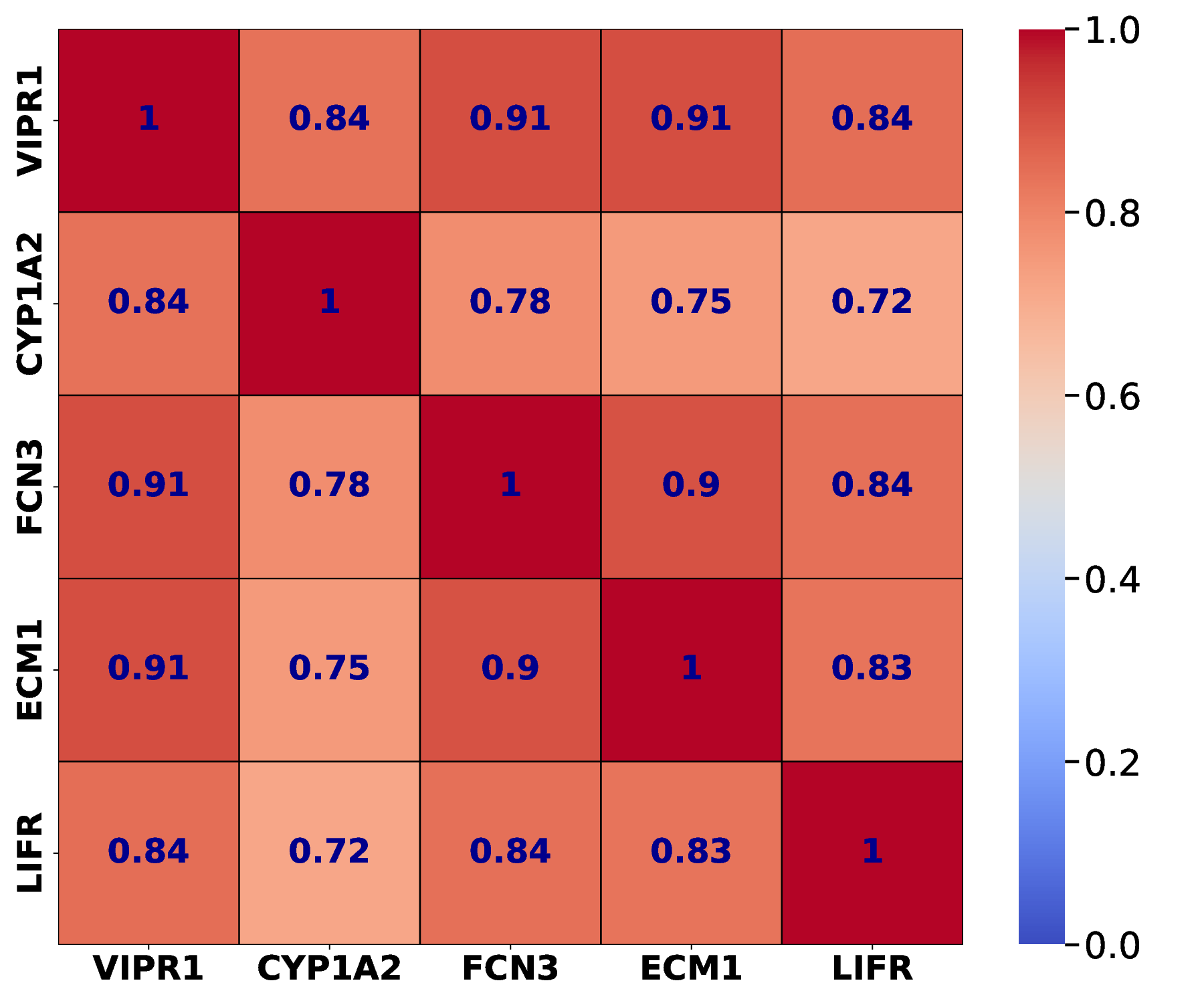}
\end{subfigure}
\hfill
\begin{subfigure}[b]{0.3\textwidth}
    \setlength{\unitlength}{1pt}
    \begin{picture}(0,0)
        \put(-10,5){\textbf{E}}
    \end{picture}
    \includegraphics[width=\textwidth]{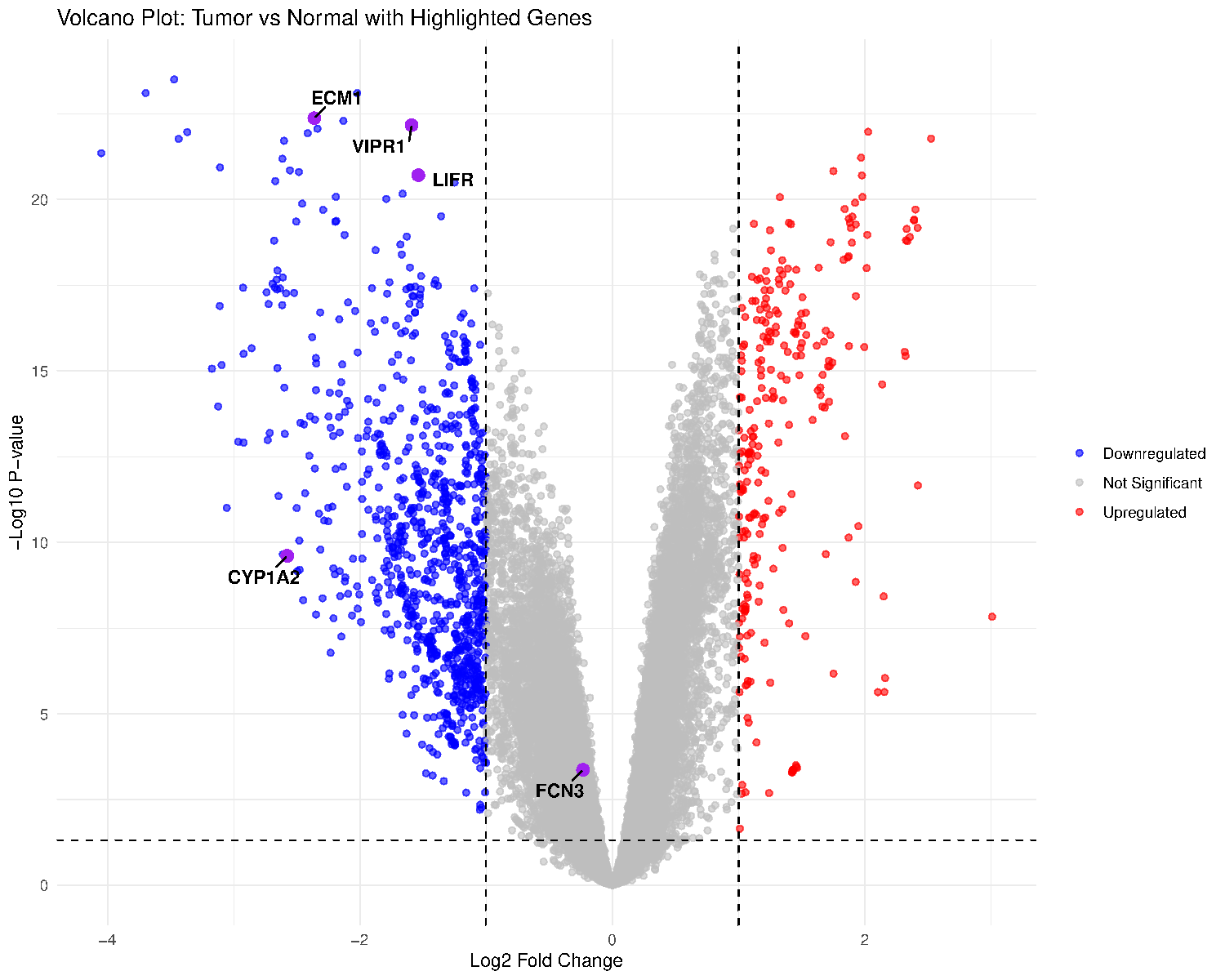}
    \includegraphics[width=\textwidth]{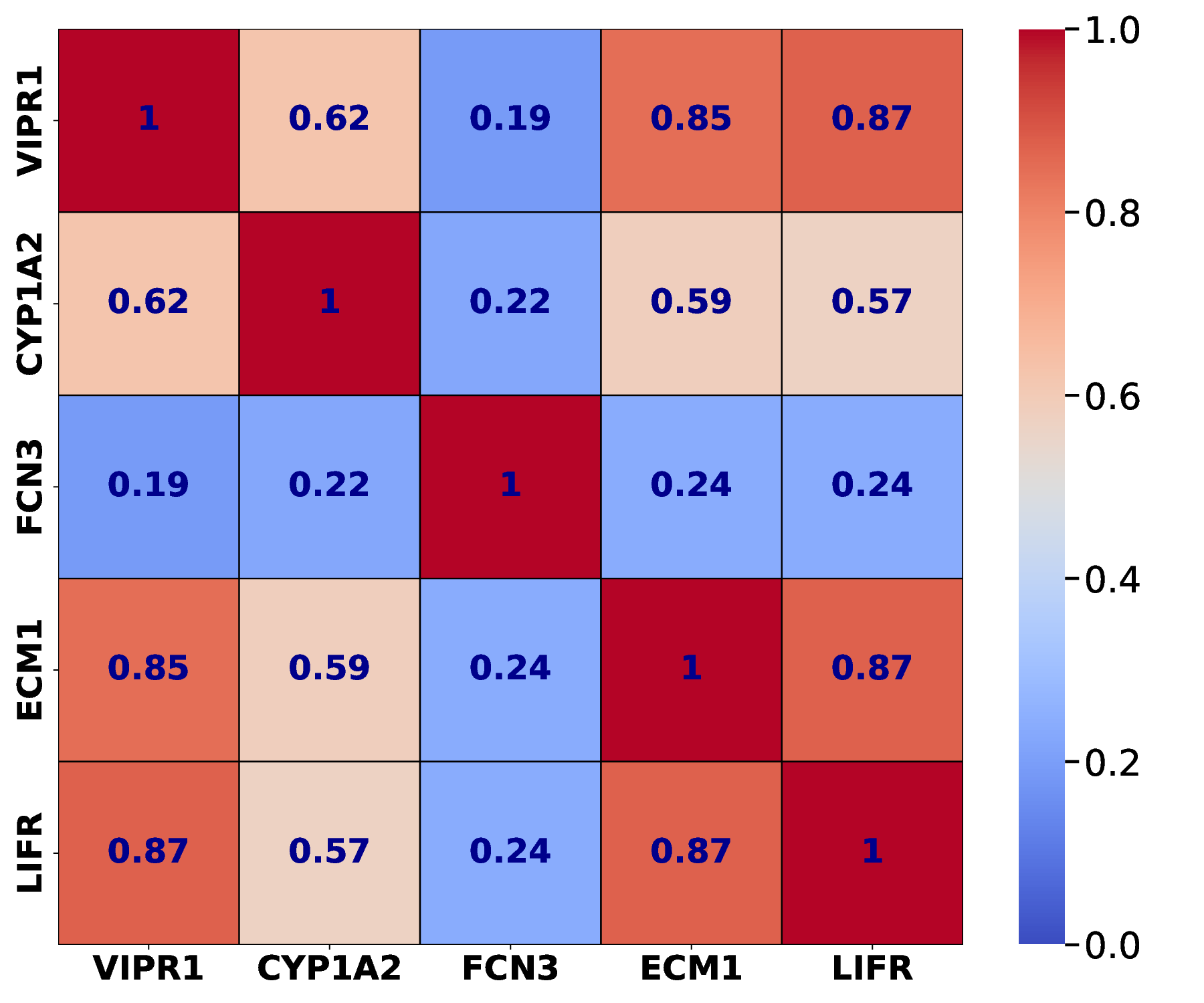}
\end{subfigure}
\hfill
\begin{subfigure}[b]{0.3\textwidth}
    \setlength{\unitlength}{1pt}
    \begin{picture}(0,0)
        \put(-10,5){\textbf{F}}
    \end{picture}
    \includegraphics[width=\textwidth]{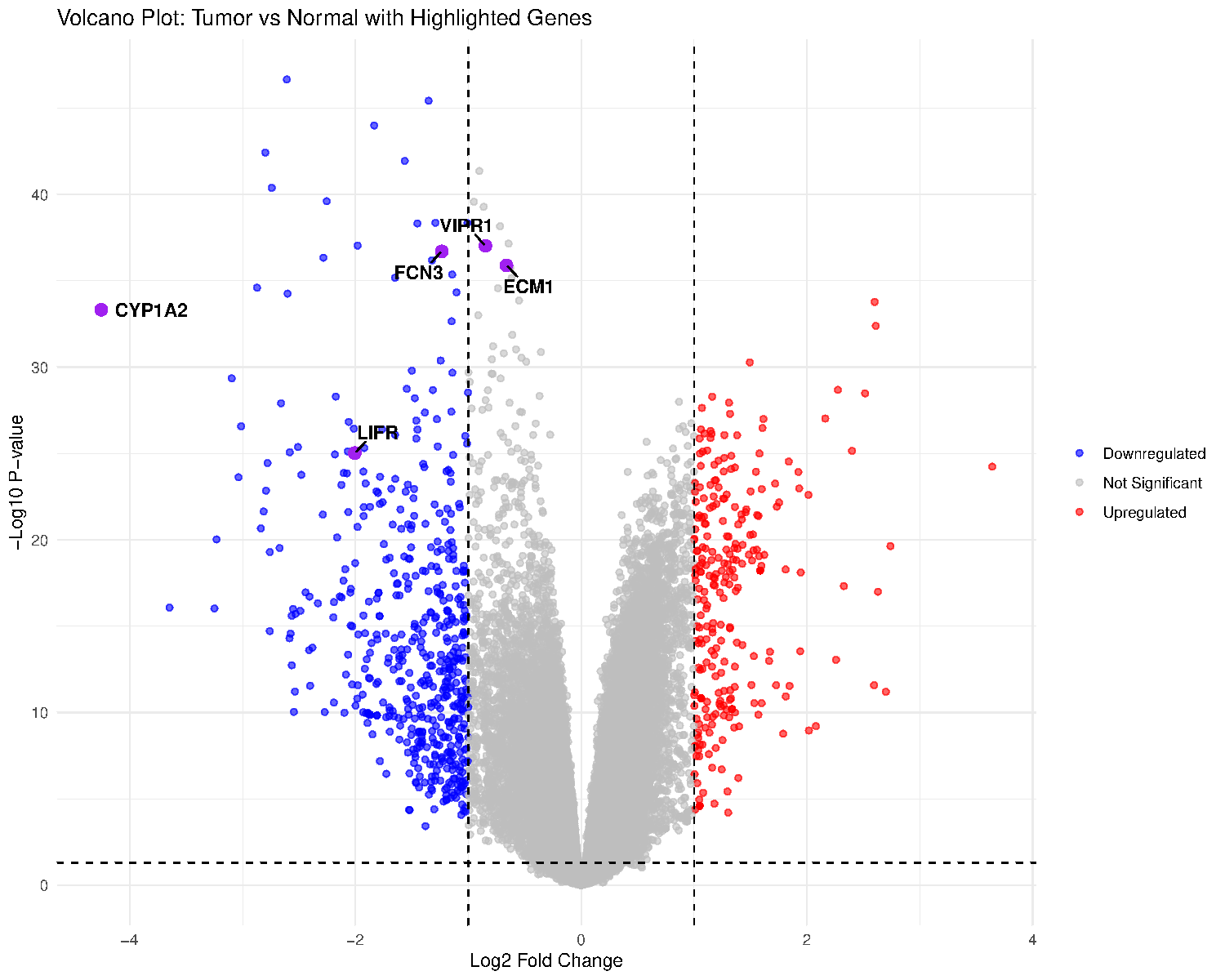}
    \includegraphics[width=\textwidth]{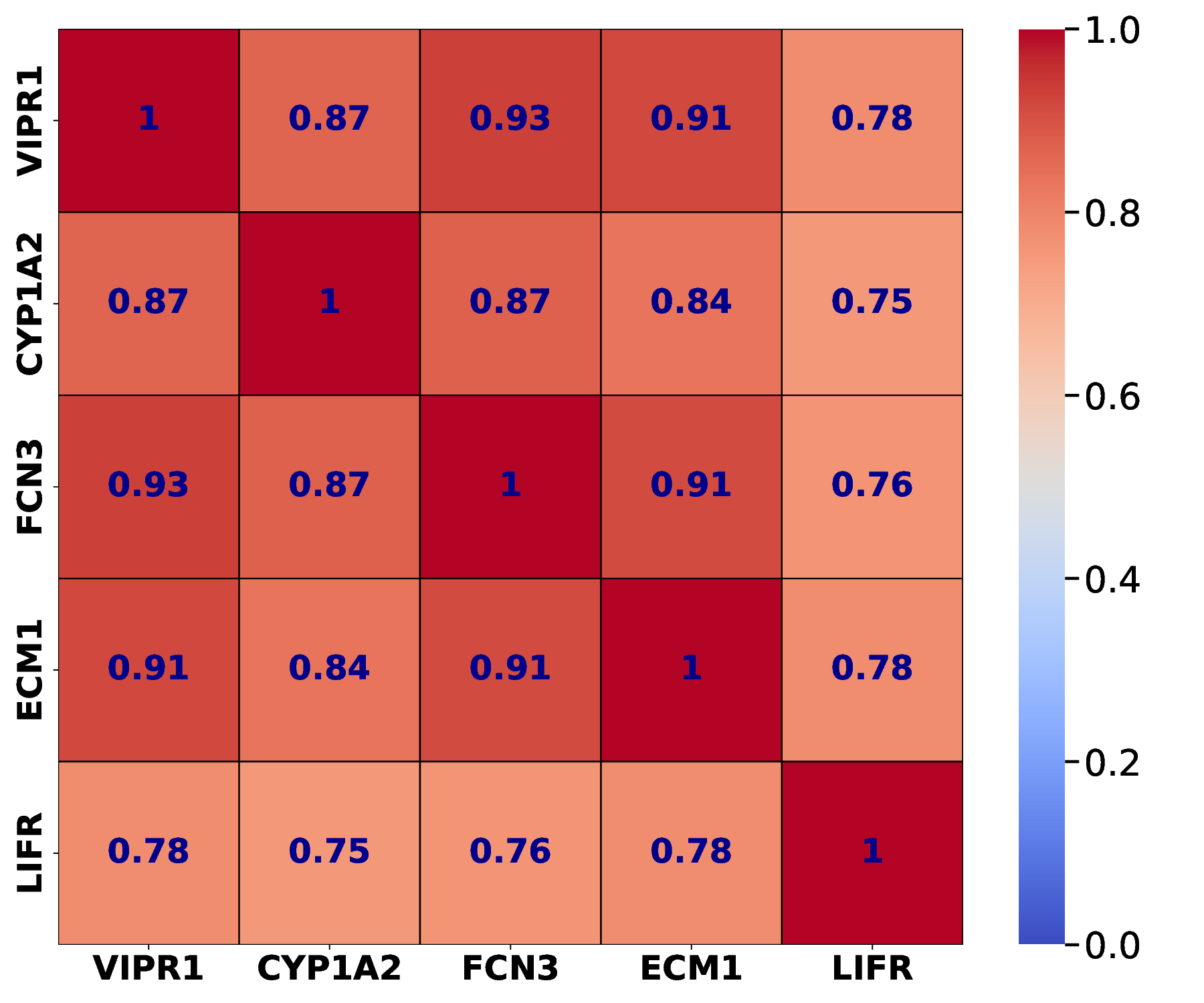}
\end{subfigure}

\caption{ Volcano plots (top) and corresponding gene correlation heatmaps (bottom) for six datasets. Panels A–F correspond to datasets GSE60502, GSE57957, GSE64041, GSE121248, GSE47197, and GSE76297, respectively. Significant genes are highlighted in the volcano plots, and correlation coefficients among selected genes are shown in the heatmaps.}
\label{fig:volccorr}
\end{figure*}

\begin{table}[]
\centering
\begin{tabular}{|c|l|}
    \hline
    \textbf{Five genes used in formula} & \textbf{Associated Drugs} \\
    \hline
    VIPR1 & Vasoactive intestinal peptide [I] \\
    \hline
    CYP1A2 &  Fluvoxamine [A,I], Caffeine [A], Anagrelide [A]\\ & Ropinirole [A,I], Theophylline[A] \\
    \hline
    FCN3 & Zinc [A,I], Zinc acetate [A,I]\\
    \hline
    ECM1 & - \\
    \hline
    LIFR & - \\
    \hline
\end{tabular}
\caption{Drugs asscociated with the five genes of VIPR1, CYP1A2, FCN3, ECM1 and LIFR. The status of each drug is shown in the brackets next to it: A and I indicate that the drug is approved and investigational, respectively.}
\label{tab:drug}
\end{table}

\end{document}